\documentstyle[12pt]{article}

\catcode`\@=11
\def\marginnote#1{}
\newcount\hour
\newcount\minute
\newtoks\amorpm
\hour=\time\divide\hour by60
\minute=\time{\multiply\hour by60 \global\advance\minute
by-\hour}\edef\standardtime{{\ifnum\hour<12
\global\amorpm={am}%
        \else\global\amorpm={pm}\advance\hour by-12 \fi
        \ifnum\hour=0 \hour=12 \fi
        \number\hour:\ifnum\minute<10
0\fi\number\minute\the\amorpm}}
\edef\militarytime{\number\hour:\ifnum\minute<10
0\fi\number\minute}

\def\draftlabel#1{{\@bsphack\if@filesw {\let\thepage\relax
   \xdef\@gtempa{\write\@auxout{\string
      \newlabel{#1}{{\@currentlabel}{\thepage}}}}}\@gtempa
   \if@nobreak \ifvmode\nobreak\fi\fi\fi\@esphack}
        \gdef\@eqnlabel{#1}}
\def\@eqnlabel{}
\def\@vacuum{}
\def\draftmarginnote#1{\marginpar{\raggedright\scriptsize\tt#1}}
\def\draft{\oddsidemargin -.5truein
        \def\@oddfoot{\sl preliminary draft \hfil
        \rm\thepage\hfil\sl\today\quad\militarytime}
        \let\@evenfoot\@oddfoot \overfullrule 3pt
        \let\label=\draftlabel
        \let\marginnote=\draftmarginnote
\def\@eqnnum{(\theequation)\rlap{\kern\marginparsep\tt\@eqnlabel}%
\global\let\@eqnlabel\@vacuum}  }

\def\numberbysection{\@addtoreset{equation}{section}
        \def\theequation{\thesection.\arabic{equation}}}
\def\underline#1{\relax\ifmmode\@@underline#1\else
 $\@@underline{\hbox{#1}}$\relax\fi}

\catcode`@=12
\relax

\numberbysection

\advance \topmargin by -\headheight
\advance \topmargin by -\headsep

\evensidemargin \oddsidemargin
\def\rf#1{(\ref{eq:#1})}
\def\lab#1{\label{eq:#1}}
\def\br{\begin{eqnarray}}
\def\er{\end{eqnarray}}
\def\be{\begin{equation}}
\def\ee{\end{equation}}
\def\nn{\nonumber}

\def\({\left(}
\def\){\right)}

\relax

\newcommand{\bi}[1]{\bibitem{#1}}
\def\PRL#1#2#3{{\sl Phys. Rev. Lett.} {\bf#1} (#2) #3}
\def\NPB#1#2#3{{\sl Nucl. Phys.} {\bf B#1} (#2) #3}

\def\CMP#1#2#3{{\sl Commun. Math. Phys.} {\bf #1} (#2) #3}
\def\PRD#1#2#3{{\sl Phys. Rev.} {\bf D#1} (#2) #3}

\def\PLB#1#2#3{{\sl Phys. Lett.} {\bf #1B} (#2) #3}
\def\JMP#1#2#3{{\sl J. Math. Phys.} {\bf #1} (#2) #3}
\def\JNMP#1#2#3{{\sl J. Nonl. Math. Phys.} {\bf #1} (#2) #3}
\def\PTP#1#2#3{{\sl Prog. Theor. Phys.} {\bf #1} (#2) #3}

\def\AoP#1#2#3{{\sl Ann. of Phys.} {\bf #1} (#2) #3}

\def\RMP#1#2#3{{\sl Rev. Mod. Phys.} {\bf #1} (#2) #3}

\def\IJMPA#1#2#3{{\sl Int. J. Mod. Phys.} {\bf A#1} (#2) #3}

\def\JPA#1#2#3{{\sl J. Physics} {\bf A#1} (#2) #3}

\def\MPLA#1#2#3{{\sl Mod. Phys. Lett.} {\bf A#1} (#2) #3}

\def\JPIV#1#2#3{{\sl J. Phys. IV} {\bf #1} (#2) #3}

\def\a{\alpha}

\def\d{\delta}
\def\D{\Delta}

\def\vp{\varphi}

\def\/{\frac}

\def\k{\kappa}
\def\l{\lambda}

\def\o{\omega}
\def\O{\Omega}

\def\pa{\partial}

\def\qq{\qquad}
\def\ra{\rightarrow}

\def\vp{\varphi}

\def\ti{\tilde}
\def\u{\upsilon}

\newcommand{\eq}{=}
\def\({\Big(}
\def\){\Big)}
\def\[{\Big[}
\def\]{\Big]}

\begin{document}
\begin{center}
{\large {\bf Generalized sine-Gordon/massive Thirring  models and soliton/particle correspondences}}
\end{center}

\normalsize

\begin{center}

Jos\'e Acosta and Harold Blas 
\par \vskip .1in \noindent
Instituto de F\'{\i}sica Te\'{o}rica\\
Universidade Estadual Paulista\\
Rua Pamplona 145,\\
01405-900\,-\,S\~{a}o Paulo, S.P.\\
Brazil

\end{center}

\begin{abstract}

We consider a real Lagrangian off-critical submodel describing the soliton sector of the so-called conformal affine $sl(3)^{(1)}$ Toda model coupled to matter fields (CATM). The theory is treated as a constrained system in the context of Faddeev-Jackiw and the symplectic schemes. We exhibit the parent Lagrangian nature of the model from which generalizations of the sine-Gordon (GSG) or the massive Thirring (GMT) models are derivable. The dual description of the model is further emphasized by providing the relationships between bilinears of GMT spinors and relevant expressions of the GSG fields. In this way we exhibit the strong/weak coupling phases and the (generalized) soliton/particle correspondences of the model. The  $sl(n)^{(1)}$ case is also outlined. 
\end{abstract}

\noindent 
PACS numbers: 11.10.Lm, 12.40.Nn, 11.30.Na, 11.27.+d\\
Keywords: (generalized) Thirring and sine-Gordon models, affine Toda coupled to matter, solitons, topological current.\\

\par \vskip 1.7in \noindent

\newpage
\newpage
\section{Introduction}
\label{sec:int}

Integrable theories in two-dimensions have been an extraordinary laboratory for the understanding of basic nonperturbative aspects of physical theories and various aspects, relevant in more realistic $4$-dimensional models, have been tested \cite{abdalla}. In particular the conformal affine Toda models coupled to (Dirac) matter fields (CATM) \cite{matter} for the $sl(2)^{(1)}$ and $sl(3)^{(1)}$ cases are discussed in \cite{bla,nucl} and \cite{bueno}, respectively. The interest in such models comes from their integrability and duality properties \cite{matter,nucl}, which can be used as toy models to understand some phenomena; such as, a confinement mechanism in QCD \cite{bla,bueno} and the electric-magnetic duality in four dimensional gauge theories, conjectured in \cite{montonen} and developed in \cite{vw}. The ATM type systems may also describe some low dimensional condensed matter phenomena, such as selftrapping of electrons into solitons, see e.g. \cite{brazovskii}, tunneling in the integer quantum Hall effect \cite{barci}, and, in particular, polyacteline molecule systems in connection with fermion number fractionization \cite{jackiw}.

Off-critical submodels, such as the $sl(2)$ affine Toda model coupled to matter fields (ATM), can be obtained at the classical or quantum mechanical level through some convenient reduction processes starting from CATM \cite{nucl,annals}. In the $sl(2)$ case, using bosonization techniques, it has been shown that the classical equivalence between the $U(1)$ vector and topological currents holds true at the quantum level, and then leads to a bag model like mechanism for the confinement of the spinor fields inside the solitons; in addition, it has been shown that the $sl(2)$ ATM theory decouples into a sine-Gordon model (SG) and a free scalar \cite{bla,witten}. These facts indicate the existence of a sort of duality in these models involving solitons and particles \cite{montonen}. The symplectic structure of the $sl(2)$ ATM model has recently been studied \cite{annals} in the context of Faddeev-Jackiw (FJ) \cite{ja} and (constrained) symplectic methods \cite{symplectic,montani}. Imposing the equivalence between the $U(1)$ vector and topological currents as a constraint there have been obtained the sine-Gordon (SG) or the massive Thirring (MT) model.

One of the difficulties with generalizations of complex affine Toda field theories, beyond $su(2)$ and its associated SG model, has to do with unitarity. Whereas, for practical applications such as low dimensional condensed matter systems (see \cite{saleursimonetti} and references therein) and $N-$body problems in nuclear physics \cite{sakamoto}, the properties of interest are usually integrability and nonperturbative results of multifield Lagrangians. Therefore, integrable quantum field theories with several fields (bosons and/or fermions) are of some importance.

In this paper we construct many field generalizations of SG/MT models based on soliton/particle duality and unitarity. Beyond the well known $sl(2)$ case the related $sl(n)^{(1)}$ CATM model does not possess a local Lagrangian, therefore we resort to an off-critical submodel Lagrangian with well behaved classical solutions making use of the results of \cite{bueno}. In \cite{bueno} the authors studied the $sl(3)^{(1)}$ CATM soliton solutions and some of their properties up to general $2$-soliton. Using the FJ and symplectic methods we show the parent Lagrangian \cite{hje} nature of the $sl(3)$ ATM model from which the generalized sine-Gordon (GSG) or the massive Thirring (GMT) models are derivable. We thus show that there are (at least classically) two equivalent descriptions of the model, by means of either the Dirac or the Toda type fields. It will also be clear the duality exchange of the coupling regimes $g \rightarrow 1/g$ and the generalized soliton/particle correspondences in each $sl(2)$ ATM submodel, which we uncover by providing explicit relationships between the GSG and GMT fields. We also outline the steps toward the $sl(n)$ affine Lie algebra generalizations. In this way we give a precise field content of both sectors; namely, the correct GMT/GSG duality, first undertaken in \cite{halpern}.

The paper is organized as follows. In section 2 we define the $sl(3)$ ATM model . Section 3 deals with the model in the FJ framework, the outcome is the GMT model. In section 4, we attack the same problem from the point of view of symplectic quantization \cite{symplectic,montani} giving the Poisson brackets of the GMT and GSG models. Section 5 deals with the soliton/particle and strong/weak coupling correspondences. Section 6 outlines the relevant steps towards the generalization to $sl(n)$ ATM.\, In the appendix \ref{app:catm} we present the construction of $sl(3)^{(1)}$ CATM model and its relationship to the (two-loop) WZNW model.

\section{Description of the model}
\label{sec:ATM}

In affine Toda type theories the question of whether all mathematical solutions are physically acceptable deserves a careful analysis, specially if any consistent quantization of the models is discussed. The requirement of real energy density leads to a certain reality conditions on the solutions of the model. In general, a few soliton solutions survive the reality constraint, if in addition one also demands positivity. These kind of issues are discussed in Refs. \cite{hermitian}. Here we follow the prescription to restrict the model to a subspace of classical solutions which satisfy the physical principles of reality of energy density and soliton/particle correspondence.

In CATM models associated to the principal gradation of an affine Lie algebra we have a $1$-soliton solution (real Toda field) for each pair of Dirac fields $\psi^{i}$\, and  $\widetilde{\psi}^{i}$ \cite{matter}. This fact allows us to make the identifications $\widetilde{\psi}^{i}\, \sim \,(\psi^{i})^{*}$, and take real Toda fields. In the case of $sl(2)^{(1)}$ CATM theory, this procedure does not spoil the particle-soliton correspondence \cite{bla,nucl}. 

We consider the $sl(3)^{(1)}$ CATM theory (see Appendix \ref{app:catm}) with the conformal symmetry gauge fixed \cite{a1} by setting $\eta\,=\,0$ and the reality conditions  
\br
\lab{real1}
\widetilde{\psi}^{j}&=&-{(\psi^{j})}^{*},\,\, (j=1,2,3);\,\,\,\,\vp_{a}^{*}= \vp_{a},\,\, (a=1,2),\\
\nn
\mbox{or}
\\
\widetilde{\psi}^{j}&=&{(\psi^{j})}^{*},\,\, j=1,2;\,\,\,\,\,\,\,\widetilde{\psi}^{3}=-{(\psi^{3})}^{*},\nn
\\\, \,\, \vp_{1,\,2}&\rightarrow& \vp_{1,\,2}- \pi \,\,\,\,\,\, (\mbox{the new}\,\vp_{a}\mbox{'s being real fields}),\lab{real2}
\er
where $*$ means complex conjugation. The condition \rf{real2} must be supplied with $x^{\mu}\rightarrow -x^{\mu}$.  
Moreover, for consistency of the equations of motion \rf{eqnm4}-\rf{eqnm15} under the reality conditions \rf{real1}-\rf{real2}, from Eqs. \rf{eqnm5}, \rf{eqnm7}, \rf{eqnm8}, \rf{eqnm10}, \rf{eqnm12} and \rf{eqnm15}, we get the relationships 
\br
\widetilde{\psi }_{L}^{j}\psi _{R}^{3}-\widetilde{\psi }_{R}^{j}\psi _{L}^{3}e^{-3i\varphi_{j}}\,=\,0,\,\,j=1,2;\,\,\,\, \psi _{L}^{1}\psi_{R}^{2}e^{-3i\varphi_{1}}-\psi _{L}^{2}\psi_{R}^{1}e^{-3i\varphi_{2}}\,=\,0. \lab{condi}
\er

Then, the above reality conditions and constraints allow us to define a suitable physical Lagrangian. The equations \rf{eqnm1}, \rf{eqnm4}-\rf{eqnm15}, supplied with \rf{real1} [or \rf{real2}] and \rf{condi}, follow from the Lagrangian
\br
\lab{atm1}
\frac{1}{k}{\cal L} = \sum_{j=1}^{3} \[ \frac{1}{24}\partial_{\mu }\phi_j \partial ^{\mu }\phi_j + i\overline{\psi}^j \gamma ^{\mu}\partial _{\mu }\psi^j - m^{j}_{\psi }\overline{\psi}^j e^{i \phi_j  \gamma_{5}}\psi^j\] 
\er
where ${\bar{\psi}}^{j} \equiv {({\psi}^{j})}^{\dagger} \,\gamma_0\,$,\, $\phi_{1}\equiv 2\vp_{1}-\vp_{2}$,\, $\phi_{2}\equiv 2\vp_{2}-\vp_{1}$,\, $\phi_{3}\equiv \phi_{1}+\phi_{2}$,\, $m_{\psi}^3=m_{\psi}^1+m_{\psi}^2$, \,$k$ is an overall coupling constant and the $\vp_{j}$ are real fields.

The Eq. \rf{atm1} defines the  $sl(3)$ {\sl affine Toda theory coupled to matter fields} (ATM). Notice that the space of solutions of $sl(3)^{(1)}$ CATM model satisfying the conditions \rf{real1}-\rf{condi} must be solutions of the $sl(3)$ ATM theory \rf{atm1}. Indeed, it is easy to verify that the three species of one-soliton solutions [$S\equiv${\sl 1-soliton}($\bar{S}\equiv${\sl 1-antisoliton})] \cite{bueno}: $\{\(\vp_{1},\, \psi^{1}\)_{S/\bar{S}},\, \vp_{2}=0,\,\psi^{2}=0,\,\psi^{3}=0\}$, $\{\(\vp_{2},\, \psi^{2}\)_{S/\bar{S}},\, \vp_{1}=0,\,\psi^{1}=0,\,\psi^{3}=0\}$ and  $\{\(\vp_{1}+\vp_{2},\,\psi^{3}\)_{S/\bar{S}},\, \vp_{1}=\vp_{2},\,\psi^{1}=0,\,\psi^{2}=0\}$\, satisfy the equations of motion; i.e., each positive root of $sl(3)$ reproduces the $sl(2)$ ATM case \cite{bla,nucl}. Moreover, these solutions satisfy the above reality conditions and constriants \rf{real1}-\rf{condi} (with \rf{real1} and \rf{real2} for $S$ and $\bar{S}$, respectively), and the equivalence between the $U(1)$ vector and topological currents \rf{equivalence}. Then, the soliton/particle correspondences survive the above reduction processes performed to define the $sl(3)$ ATM theory. 

The class of $2$-soliton solutions of $sl(3)^{(1)}$ CATM \cite{bueno} behave as  follows:\, i) they are given by 6 species associated to the pair $(\a_{i},\a_{j}),\, i\le j;\,\, i,j=1,2,3$; where the $\a$'s are the positive roots of $sl(3)$ Lie algebra. Each species $(\a_{i},\a_{i})$ solves the $sl(2)$ CATM submodel\cite{a2}; 
ii) satisfy the $U(1)$ vector and topological currents equivalence \rf{equivalence}.

\section{The generalized massive Thirring model (GMT)}
\label{sec:FJ}

Let us consider the following Lagrangian
\be
\lab{lagrangian1}
       \frac{1}{k} {\cal L} = \sum_{j=1}^{3}\Big[\frac{1}{24}\partial_{\mu }\phi_j \partial ^{\mu }\phi_j 
        + i\overline{\psi}^j \gamma ^{\mu}\partial _{\mu }\psi^j - m^{j}_{\psi }\overline{\psi}^j e^{i \phi_j  \gamma_{5}}\psi^j 
+ \lambda^{j}_{\mu} ( m^{j } \overline{\psi}^{j}\gamma^{\mu}\psi^{j} - \epsilon^{\mu\nu}\partial_{\nu} (q_j \phi_j))\Big], 
\ee
where the ATM Lagrangian \rf{atm1} is supplied with the constraints,  
$ ( m^{l} \overline{\psi}^{l}\gamma^{\mu}\psi^{l}+\frac{m^{3}}{2} \overline{\psi}^{3}\gamma^{\mu}\psi^{3} - \epsilon^{\mu\nu}\partial_{\nu} \phi_{l})$, ($l=1,2$), with the help of the Lagrange multipliers 
$\lambda^{j}_{\mu}$\, ($\lambda^{3}_{\mu} \equiv \frac{\lambda^{1}_{\mu}+\lambda^{2}_{\mu}}{2} $,\, $q_{1}\equiv q_{2} \equiv 1$,\, $q_{3}\equiv 0$). Their total sum bears an intriguing resemblance to the  $U(1)$ vector and topological currents equivalence \rf{equivalence}; however, the $m^{j}$'s here are some arbitrary parameters. The same procedure has been used, for example, to incorporate the left-moving condition in the study of chiral bosons in two dimensions \cite{siegel}. The constraints in \rf{lagrangian1} will break the left-right local symmetries \rf{leri1}-\rf{leri2} of $sl(3)$ ATM \rf{atm1}. In order to apply the Faddeev-Jackiw (FJ) method we should write \rf{lagrangian1} in the first order form in time derivative, so let us define the conjugated momenta 
\br
\lab{moments}
\pi _{1}&\equiv& \pi_{\phi_{1}}\,=\,\frac{1}{12} ( 2 \dot{\phi}_1 + \dot{\phi}_2) + \lambda^{1}_{1},
        \qquad 
\pi _{2} \equiv \pi_{\phi_{2}}\,=\,\frac{1}{12} ( 2 \dot{\phi}_2 + \dot{\phi}_1) + \lambda^{2}_{1} ,   \nonumber
        \\
\pi _{\lambda^{1}_{\mu}}&=& 0,       
        \qquad                                  
\pi _{\lambda^{2}_{\mu}} = 0,                      
        \qquad 
\pi^{j}_{R}\equiv \pi^{j}_{\psi_{R}} =-i{\ti \psi}^{j}_{R},
        \qquad 
\pi^{j}_{L} \equiv \pi^{j}_{\psi_{L}} =-i{\ti \psi}^{j}_{L}.
\er

We are assuming that Dirac fields are anticommuting Grasmannian variables and their momenta variables defined through {\bf left} derivatives.   
Then, as usual, the Hamiltonian is defined by (sum over repeated indices is assumed)
\be
\lab{hamiltonian}
        {\cal H}_{c} = \pi_{1} \dot{\phi}_{1} + \pi_{2} \dot{\phi}_{2} 
+ \dot{\psi}^{j}_{R} \pi^{j}_{R}  +  \dot{\psi}^{j}_{L} \pi^{j}_{L}     - \cal{L}.
\ee

Explicitly the Hamiltonian density becomes 
\begin{eqnarray}
\lab{hamiltonian1}
{\mathcal{H}}_{c} &=& 2 ( \pi_{j})^{2} + 4 ( \lambda^{1}_{1})^{2} 
+ 4\left ( \lambda^{2}_{1}\right)^{2} - \lambda^{1}_{1} 
{\mathcal{J}}_{1} - \lambda^{2}_{1} {\mathcal{J}}_{2} - 4 \left( \lambda^{1}_{1} \lambda^{2}_{1} \right)  \nonumber 
        \\
&&+\frac{1}{24}\left( \phi _{j,x}\right) ^{2}-\pi _{R}^{j}\psi_{R,x}^{j} + 
        \pi _{L}^{j}\psi _{L,x}^{j}+im_{\psi }^{j}(e^{-\phi _{j}}
        \tilde{\psi}_{R}^{j}\psi _{L}^{j}-e^{\phi _{j}}\tilde{\psi}_{L}^{j}\psi _{R}^{j}) 
\nonumber
\\
&&+\lambda^{1}_{0}[J_{1}^{0}-\phi _{1,x}] 
+\lambda^{2}_{0}[J_{2}^{0}-\phi _{2,x}],  
\end{eqnarray}
where $\pi _{3} \equiv \pi _{1}-\pi _{2}$,\, ${\mathcal{J}}_{1} \equiv J_{1}^{1}+ 4\left( 2\pi_{1}-\pi _{2}\right)$,\,\,
        ${\mathcal{J}}_{2} \equiv J_{2}^{1} 
        + 4\left( 2\pi _{2}-\pi _{1}\right)$\, 
 and
\br
J_{1}^{\mu}\,=\,m^{1} j^{\mu}_{l}+\frac{m^{3}}{2} j^{\mu}_{3};\,\,\,\,J_{2}^{\mu}\,=\,m^{2} j^{\mu}_{2}+\frac{m^{3}}{2} j^{\mu}_{3};\,\,\,\,\,\,j^{\mu}_{l}\,\equiv\, \bar{\psi}^{l}\gamma^{\mu}\psi^{l},\,\,\,\,l=1,2,3.
\er

Let us observe that each $U(1)$ Noether current of the $sl(3)$ ATM theory defined in \rf{atm1} is conserved separately; i.e., $\pa_{\mu}j^{\mu}_{l}=0,\,\,\,\, l=1,2,3 $.  

Next, the same Legendre transform \rf{hamiltonian} is used to write the first order Lagrangian 
\br
\lab{lagran}
        {\cal L}& = \pi_{1}\dot{\phi}_{1}  + \pi_{2}\dot{\phi}_{2} 
        +  \dot{\psi}^{j}_{R} \pi^{j}_{R}  + \dot{\psi}^{j}_{L} \pi^{j}_{L} 
        - {\cal H}_{c}.
\er

Our starting point for the FJ analysis will be this first order Lagrangian. Then the Euler-Lagrange equations for the Lagrange multipliers allow one to solve two of them 
\br
\lambda^{1}_{1}= \frac{2 {\mathcal{J}}_{1} +  {\mathcal{J}}_{2} }
                {12 }, \qq                                
\lambda^{2}_{1}=  \frac{2  {\mathcal{J}}_{2} +  {\mathcal{J}}_{1} }
                {12 }
\er
and the remaining equations lead to two constraints
\be
\lab{cons}
        \O_{1} \equiv J_{1}^{0}-\phi _{1,x} = 0, 
        \qquad
        \O_{2} \equiv J_{2}^{0}-\phi _{2,x} = 0.
\ee

The Lagrange multipliers $\lambda^{1}_{1}$ and $\lambda^{2}_{1}$ must be 
replaced back in \rf{lagran} and the constraints \rf{cons} solved. Firstly, 
let us replace the $\lambda^{1}_{1}$ and $\lambda^{2}_{1}$ multipliers into 
${\cal H}_{c}$, then one gets
\br
\lab{hamiltonian2}
{\cal H}_{c}^{\prime} &=& 2(\pi_{j})^{2} - \frac{1}{12}
        \{ ( {\cal J}_1)^2 + ( {\cal J}_2)^2 
        +({\cal J}_1 {\cal J}_2)\} + \frac{1}{24} ( \phi_{j,x})^2
        \nn
        \\          
        &+& i \tilde{\psi}^{j}_{R} {\psi}^{j}_{R,x} 
         -  i \tilde{\psi}^{j}_{L} {\psi}^{j}_{L,x}
        + im^{j}_{\psi} ( e^{-i\phi_{j}} \tilde{\psi}^{j}_{R}\psi^{j}_{L} 
                        - e^{ i\phi_{j}} \tilde{\psi}^{j}_{L}\psi^{j}_{R}). 
\er

The new Lagrangian becomes
\be
\lab{lagran1}
        {\cal L}^{\prime} = \pi_{1}\dot{\phi}_1 + \pi_{2}\dot{\phi}_2 
        + \dot{\psi}^{j}_{R} \pi^{j}_{R} 
        + \dot{\psi}^{j}_{L} \pi^{j}_{L} - {\cal H}_{c}^{\prime}.
\ee

We implement the constraints \rf{cons} by replacing in \rf{lagran1} the fields 
$\phi_{1},\,\phi_{2}$ in terms of the space integral of the current components $J^{0}_{1},\,J^{0}_{2}$. Then we get the Lagrangian 
\begin{eqnarray}
\lab{lagran2}
{\cal L}^{\prime\prime} &=& \pi_{1} \pa_{t} \int^{x} J^{0}_{1} 
                                  + \pi_{2} \pa_{t} \int^{x} J^{0}_{2} 
                                  + \dot{\psi}^{j}_{R} \pi^{j}_{R} 
                                + \dot{\psi}^{j}_{L} \pi^{j}_{L}  
         - i {\tilde \psi}^{j}_{R} \psi_{R,x}^{j} + i{\tilde \psi}^{j}_{L} {\psi}_{L,x}^{j} 
         \\                                                                                             \nonumber
         &-& i m^{j}_{\psi} \(  e^{-i \int^{x} J^{0}_{j}} {\tilde \psi}^{j}_{R} \psi^{j}_{L}
                            - e^{ i \int^{x} J^{0}_{j}} {\tilde \psi}^{j}_{L} \psi^{j}_{R}  \) 
         - \frac{1}{12}  \((J_{1})^2 + (J_{2})^2 
         + J_{1}.J_{2} \)
         \\                                                                                             \nn
         &+& \pi_1 J^{1}_{1} +  \pi_2 J^{1}_{2},
\end{eqnarray}
where $J^{0}_{3} \equiv J^{0}_{1}+J^{0}_{2}$. Observe that the terms containig the $\pi_{a}$'s in Eq. \rf{lagran2} cancel to each other if one uses the current conservation laws. Notice the apperances of various types of current-current interactions. The following Darboux transformation
\br
\lab{darboux}
\psi^{j}_{R} \rightarrow e^{-\frac{i}{2} \int^{x} J^{0}_{j}} \psi^{j}_{R}, 
        \qquad  
        \psi^{j}_{L} \rightarrow e^{    \frac{i}{2} \int^{x} J^{0}_{j}} \psi^{j}_{L},\,\,\,\,\,\,j=1,2,3,
\er
is used to diagonalize the canonical one-form. Then, the kinetic terms will give additional current-current interactions, $-\frac{1}{2}[J_{1}.(j_{1}+j_{3})+ J_{2}.(j_{2}+j_{3})]$.  We are, thus, after defining $k\equiv 1/g$, and rescaling the fields $\psi^{j} \rightarrow 1/\sqrt{k}\, \psi^{j}$, left with the Lagrangian 
\begin{equation}
\lab{thirring1}
{\cal L}[\psi,\overline{\psi}]= \sum_{j=1}^{3}\{i\overline{\psi}^{j}\gamma^{\mu}\pa_{\mu}\psi^{j} 
        + m^{j}_{\psi}\,\,{\overline{\psi}}^{j}\psi^{j}\}\,
        \,- \sum_{{\tiny\begin{array}{c} k,l=1\\k\leq l\end{array}}}^{3}\[\bar{a}_{kl}j_{k}.j_{l}\], 
\end{equation}
where $\bar{a}_{kl}=g\, a_{kl}$, with $a_{33}=\frac{1}{2}(\frac{(m^{3})^2}{8}+m^{3})$; $a_{12}=\frac{1}{12}m^{1}m^{2}$; $a_{ii}=\frac{1}{2}(\frac{(m^{i})^2}{6}+m^{i})$; $a_{i3}=\frac{1}{2}(\frac{m^{i} m^{3}}{4}+m^{i}+ \frac{m^{3}}{2})$, $i=1,2$. This defines the {\sl generalized massive Thirring model} (GMT). The canonical pairs are  $(-i{\tilde \psi}_{R}^{j}, \psi_{R}^{j})$ and $(-i{\tilde \psi}_{L}^{j}, \psi_{L}^{j})$.

\section{The symplectic formalism and the ATM model}

\subsection{The (constrained) symplectic formalism}

\label{sec:symplectic1}

We give a brief overview of the basic notations of symplectic approach \cite{a3}. The geometric structure is defined by the closed (pre)symplectic two-form
\br
f^{(0)}\,=\,\frac{1}{2}f^{(0)}_{ij}(\xi^{(0)}) d{\xi^{(0)}}^{i}\wedge d{\xi^{(0)}}^{j}
\er
where
\br
\lab{form}
f^{(0)}_{ij}(\xi^{(0)})\eq\frac{\pa}{\pa{{\xi}^{(0)}}^{i}}{\bf a}^{(0)}_{j}(\xi^{(0)})-\frac{\pa}{\pa{{\xi}^{(0)}}^{j}}{\bf a}^{(0)}_{i}(\xi^{(0)})
\er 
with ${\bf a}^{(0)}(\xi^{(0)})={\bf a}^{(0)}_{j}(\xi^{(0)}) d{\xi^{(0)}}^{j}$ being the canonical one-form defined from the original first order Lagrangian
\br
L^{(0)}dt\,=\, {\bf a}^{(0)}(\xi^{(0)})-V^{(0)}(\xi^{(0)})dt.
\er

The superscript $(0)$ refers to the original Lagrangian, and is indicative of the iterative nature of the computations. The constraints are imposed through Lagrange multipliers which are velocities, and in such case one has to extend the configuration space \cite{symplectic,montani}. The corresponding Lagrangian gets modified and consequently the superscript also changes. The algorithm terminates once the symplectic matrix turns out to be non-singular.

\subsection{The generalized massive Thirring model (GMT)}

\label{sec:symplectic2}

Next, we will consider our model in the framework of the symplectic formalism. Let ${\cal L}^{\prime}$, Eq. \rf{lagran1}, be the zeroth-iterated Lagrangian ${\cal L}^{(0)}$. Then the first iterated lagrangian will be 
\be
\lab{itera}
        {\cal L}^{(1)} = \pi_{1} {\dot \phi}_{1} + \pi_{2} {\dot \phi}_{2} 
        + \dot{\psi}^{j}_{R} \pi^{j}_{R} + \dot{\psi}^{j}_{L} \pi^{j}_{L} 
        + \dot{\eta}^{1}\O_{1} + \dot{\eta}^{2}\O_{2} - {\cal V}^{(1)},
\ee
where the once-iterated symplectic potential is defined by 
\be
{\cal V}^{(1)}={\cal H}_{c}^{\prime}|_{\O_{1}=\O_{2}=0}, 
\ee
and the stability conditions of the symplectic constraints, $\O_{1}$ and $\O_2$, under 
        time evolution have been implemented by making $\l^{1}_{0} \rightarrow \dot{\eta}^{1}$ and
        $\l^{2}_{0} \rightarrow \dot{\eta}^{2}$. Consider the once-iterated set of symplectic 
        variables in the following order 
\be
{\bf \xi^{(1)}} = (\eta^{1}, \eta^{2}, \phi_1, \phi_2, \psi^{1}_{R}, \psi^{1}_{L}, \psi^{2}_{R},
        \psi^{2}_{L}, \psi^{3}_{R}, \psi^{3}_{L}, \pi_{1}, \pi_2, \pi^{1}_{R}, \pi^{1}_{L},
        \pi^{2}_{R}, \pi^{2}_{L}, \pi^{3}_{R}, \pi^{3}_{L} ),
\ee
and the components of the canonical one-form
\be
{\bf a^{(1)}} = (\O_{1}, \O_2, \pi_{1}, \pi_2, -\pi^{1}_{R}, -\pi^{1}_{L}, -\pi^{2}_{R},
        -\pi^{2}_{L}, -\pi^{3}_{R}, -\pi^{3}_{L}, 0, 0, 0, 0, 0, 0, 0, 0).
\ee

These result in the singular symplectic two-form $18\mbox{x}18$ matrix
\br
f^{(1)}_{AB}(x,y)&=& \left(\begin{array}{cc}
        a_{11} &  a_{12}                        \\
        a_{21} &  a_{22}                        
        \end{array}\right)\d(x-y),
\er
where the $9\mbox{x}9$ matrices are 
\br
{\tiny
a_{11} = \left(\begin{array}{cccccccccc}
        0 & 0 &  \pa_{x} & 0 & i m^1 {\tilde \psi}^{1}_{R}  
                                      & i m^1 {\tilde \psi}^{1}_{L} 
        & 0 & 0 & \frac{im^{3}}{2} {\tilde \psi}^{3}_{R}
\\
        0 & 0 & 0 & \pa_{x} & 0 & 0 & i m^2 {\tilde \psi}^{2}_{R}  &  
        i m^2 {\tilde \psi}^{2}_{L}  & \frac{im^{3}}{2} {\tilde \psi}^{3}_{R}               
        \\
         \pa_x &  0 & 0 & 0 & 0 & 0 & 0 & 0 & 0 
        \\
        0 &  \pa_x & 0 &  0 & 0 & 0 & 0 & 0 & 0 
        \\
        i m^1 {\tilde \psi}^{1}_{R} & 0 & 0 & 0 & 0 & 0 & 0 & 0 & 0
        \\
        i m^1 {\tilde \psi}^{1}_{L} & 0 &  0 & 0 & 0 & 0 & 0 & 0 & 0
        \\
        0 & i m^2 {\tilde \psi}^{2}_{R} & 0 & 0 & 0 & 0 & 0 & 0 & 0
        \\
        0 & i m^2 {\tilde \psi}^{2}_{L} & 0 & 0 & 0 & 0 & 0 & 0 & 0
        \\
        \frac{im^{3}}{2} {\tilde \psi}^{3}_{R} & 
        \frac{im^{3}}{2} {\tilde \psi}^{3}_{R} & 0 & 0 & 0 & 0 & 0 & 0 & 0
        \end{array}\right)}, \nn
\er
\br     {\tiny
a_{12} = \left(\begin{array}{cccccccccc}
         \frac{im^{3}}{2} {\tilde \psi}^{3}_{L} & 0 & 0 
         & m^1 {\psi}^{1}_{R} 
         
& m^1 {\psi}^{1}_{L} 
         & 0 & 0 & \frac{m^{3}}{2} {\psi}^{3}_{R} &
                   \frac{m^{3}}{2} {\psi}^{3}_{L}                
        \\
        \frac{im^{3}}{2} {\tilde \psi}^{3}_{L} & 0 & 0 & 0 & 0 & 
        m^2 { \psi}^{2}_{R} & m^2 {\psi}^{2}_{L} &
        \frac{m^{3}}{2} { \psi}^{3}_{R} &
        \frac{m^{3}}{2} { \psi}^{3}_{L}          
        \\
        0 & -1 & 0 & 0 & 0 & 0 & 0 & 0 & 0 
        \\
        \vdots & 0 & \ddots & \vdots & \vdots & \vdots & \vdots & \vdots & \vdots\\
0 & 0 & 0 & 0 & 0 & 0 & 0 & -1 & 0
        \end{array}\right)}, \nn
\er
\br 
{\tiny
a_{21} = \left(\begin{array}{ccccccccc}
         \frac{im^{3}}{2} {\tilde \psi}^{3}_{L} & 
         \frac{im^{3}}{2} {\tilde \psi}^{3}_{L} & 0 & 0 & 0 & 0 & 0 & 0 & 0             
        \\
        0 & 0 & 1 & 0 & 0 & 0 & 0 & 0 & 0
        \\
        0 & 0 & 0 & 1 & 0 & 0 & 0 & 0 & 0 
        \\
        m^1 {\psi}^{1}_{R} & 0 & 0 & 0 & -1 & 0 & 0 & 0 & 0
        \\
        m^1 {\psi}^{1}_{L} & 0 & 0 & 0 & 0 & -1 & 0  & 0 & 0
        \\
        0 & m^2 {\psi}^{2}_{R} & 0 & 0 & 0 & 0 & -1  & 0 & 0
        \\
        0 & m^2 {\psi}^{2}_{L} & 0 & 0 & 0 & 0 & 0 & -1 & 0
        \\
        \frac{m^{3}}{2} {\psi}^{3}_{R} & 
        \frac{m^{3}}{2} {\psi}^{3}_{R} & 0 & 0 & 0 & 0 & 0 & 0 & -1
        \\
        \frac{m^{3}}{2} {\psi}^{3}_{L} & 
        \frac{m^{3}}{2} {\psi}^{3}_{L} & 0 & 0 & 0 & 0 & 0 & 0 & 0
        \end{array}\right)}, \,\,
{\tiny
a_{22} = \left(\begin{array}{ccccc}
        0 & 0 & \cdots & 0 & -1\\
        0 & 0 & 0 & 0 & 0\\
        \multicolumn{5}{c}\dotfill 
        \\
        0 & 0 & 0 & 0 & 0 
        \\
        -1 & 0 & \cdots & 0 & 0
        \end{array}\right)}. \nn
\er

This matrix has the zero modes
\br
        {\bf {{v}}^{(1)}}^{T}(x) &=& \( \frac{-u}{m^1}, \; \frac{-\u}{m^2}, \; 
        0, \; 0,  \; u\psi^1_{R}, \; u\psi^1_{L}, \; \u \psi^2_{R}, \; \u \psi^2_{L}, \; 
        \frac{m^3}{2} \(\frac{u}{m^1} + \frac{\u}{m^2}\) \psi^3_{R},      \nn
        \\
        && \frac{m^3}{2} \(\frac{u}{m^1} + \frac{\u}{m^2}\) \psi^3_{L}, \;
        -\frac{u^{\prime}}{m^1}, \; -\frac{\u^{\prime}}{m^2}, \;
        i u{\tilde \psi}^1_{R}, \; i u{\tilde \psi}^1_{L}, \;
        i\u{\tilde \psi}^2_{R}, \; i\u{\tilde \psi}^2_{L}, \;                   \nn
        \\
        && i\frac{m^3}{2} (\frac{u}{m^1} + \frac{\u}{m^2}) {\tilde \psi}^3_{R}, \;
        \frac{im^3}{2} (\frac{u}{m^1} + \frac{\u}{m^2}) {\tilde \psi}^3_{L} \),
\er
where $u$ and $\u$ are arbitrary functions. The zero-mode condition gives
\be
        \int {dx {\bf{{v}}^{(1)}}^T(x)\frac{\d}{\d \xi^{(1)}(x)}\int{dy{\cal V}^{(1)}}}\equiv 0.
\ee

Thus, the gradient of the symplectic potential happens to be orthogonal to the zero-mode 
${\bf {v}}^{(1)}$. Since the equations of motion are automatically validated no symplectic
constraints appear. This happens due to the presence of the symmetries of the action  
\begin{eqnarray}
\d \xi^{(1)}_{A}\,=\, {\bf v^{(1)}_{A}}(x); \,\,\,\,A=1,2,...18.     
\end{eqnarray}

So, in order to deform the symplectic matrix into an invertible one, we have to add some gauge fixing terms to the symplectic potential. One can choose any consitent set of gauge fixing 
conditions \cite{montani}. In our case we have two symmetry generators associated to the 
parameters $u$ and $v$, so there must be two gauge conditions. Let us choose
\be
\lab{gauge1}
        \O_{3}\equiv \phi_{1} = 0  \qq  \O_{4}\equiv \phi_{2} = 0.
\ee

These conditions gauge away the fields $\phi_{1}$ and $\phi_{2}$, so only the remaining field variables 
will describe the dynamics of the system. Other gauge conditions, which eventually gauge away the spinor fields
$\psi^{i}$ will be considered in the next subsection.
 
Implementing the consistency conditions by means of Lagrange multipliers $\eta^{3}$ and $\eta^{4}$ we get the twice-iterated Lagrangian
\be
        {\cal L}^{(2)} = \pi_{1}\dot{\phi}_{1} + \pi_{2}\dot{\phi}_{2} + \dot{\psi}_{R}\pi^{j}_{R}
        + \dot{\psi}_{L}\pi^{j}_{L}  + \dot{\eta}^{1}\O_{1} + \dot{\eta}^{2}\O_{2} 
        + \dot{\eta}^{3}\O_{3} + \dot{\eta}^{4}\O_{4} - {\cal V}^{(2)},
\ee
where
\br
        {\cal V}^{(2)}\;&=&\;{\cal V}^{(1)}|_{\O_{3} = \O_{4} = 0}.                      \nn
\er

Assuming now that the new set of symplectic variables is given in the following order 
\be
{\bf \xi^{(2)}} = (\eta^{1}, \eta^{2},\eta^{3}, \eta^{4}, \phi_1, \phi_2, \psi^{1}_{R}, \psi^{1}_{L}, 
        \psi^{2}_{R}, \psi^{2}_{L}, \psi^{3}_{R}, \psi^{3}_{L}, \pi_{1}, \pi_2, \pi^{1}_{R}, 
        \pi^{1}_{L},
\pi^{2}_{R}, \pi^{2}_{L}, \pi^{3}_{R}, \pi^{3}_{L} ),
\ee
and the non vanishing components of the canonical one-form
\be
{\bf a^{(2)}} = (\O_1, \O_2, \O_3, \O_4, \pi_{1}, \pi_2, -\pi^{1}_{R}, -\pi^{1}_{L}, -\pi^{2}_{R},
        -\pi^{2}_{L}, -\pi^{3}_{R}, -\pi^{3}_{L}, 0, 0, 0, 0, 0, 0, 0, 0),
\ee
one obtains the singular twice-iterated symplectic $20$x$20$ matrix 
\br
f^{(2)}_{AB}(x,y) &=& \left(\begin{array}{cc}
        a_{11} &  a_{12}                        \\
        a_{21} &  a_{22}                        
        \end{array} \right) \d(x-y),
\er
where the $10$x$10$ matrices are
\br     {\tiny
a_{11} = \left(\begin{array}{cccccccccc}
        0 & 0 &0 & 0 &  \pa_{x} & 0 & i m^1 {\tilde \psi}^{1}_{R}  
        & i m^1 {\tilde \psi}^{1}_{L} & 0 & 0              
        \\
        0 & 0 & 0 & 0 & 0 & \pa_{x} & 0 & 0 & i m^2 {\tilde \psi}^{2}_{R}  &  
        i m^2 {\tilde \psi}^{2}_{L}                
        \\
        0 & 0 & 0 & 0 & -1 & 0 & 0 & 0 & 0 & 0 
        \\
        0 & 0 & 0 & 0 & 0 & -1 & 0 & 0 & 0 & 0  
        \\
         \pa_x & 0 & 1 & 0 & 0 & 0 & 0 & 0 & 0 & 0 
        \\
        0 &  \pa_x & 0 & 1 & 0 & 0 & 0 & 0 & 0 & 0 
        \\
        i m^1 {\tilde \psi}^{1}_{R} & 0 & 0 & 0 & 0 & 0 & 0 & 0 & 0 & 0
        \\
        i m^1 {\tilde \psi}^{1}_{L} & 0 & 0 & 0 & 0 & 0 & 0 & 0 & 0 & 0
        \\
        0 & i m^2 {\tilde \psi}^{2}_{R} & 0 & 0 & 0 & 0 & 0 & 0 & 0 
& 0
        \\
        0 & i m^2 {\tilde \psi}^{2}_{L} & 0 & 0 & 0 & 0 & 0 & 0 & 0
& 0
        \end{array}\right)}, \nn
\er
\br
{\tiny
a_{12} = \left(\begin{array}{cccccccccc}
            \frac{im^{3}}{2} {\tilde \psi}^{3}_{R}  
          & \frac{im^{3}}{2} {\tilde \psi}^{3}_{L} & 0 & 0 & m^1 {\psi}^{1}_{R}  
                                                               & m^1 {\psi}^{1}_{L} 
         & 0 & 0 & \frac{m^{3}}{2} {\psi}^{3}_{R} &
                   \frac{m^{3}}{2} {\psi}^{3}_{L}                
        \\
        \frac{im^{3}}{2} {\tilde \psi}^{3}_{R} & 
        \frac{im^{3}}{2} {\tilde \psi}^{3}_{L} & 0 & 0 & 0 & 0 
        & m^2 {\psi}^{2}_{R} 
        & m^2 {\psi}^{2}_{L} 
        &\frac{m^{3}}{2} { \psi}^{3}_{R} 
        &\frac{m^{3}}{2} { \psi}^{3}_{L}                 
        \\
        0 & 0 & 0 & 0 & 0 & 0 & 0 & 0 & 0 & 0
        \\
        0 & 0 & 0 & 0 & 0 & 0 & 0 & 0 & 0 & 0
        \\
        0 & 0 & -1 & 0 & 0 & 0 & 0 & 0 & 0 & 0 
        \\
        0 & 0 & 0 & -1 & 0 & 0 & 0 & 0 & 0 & 0 
        \\
        0 & 0 & 0 & 0 & -1 & 0 & 0 & 0 & 0 & 0
        \\
        0 & 0 & 0 & 0 & 0 & -1 & 0 & 0 & 0 & 0
        \\
        0 & 0 & 0 & 0 & 0 & 0 & -1 & 0 & 0 & 0
        \\
        0 & 0 & 0 & 0 & 0 & 0 & 0 & -1 & 0 & 0
        \end{array}\right)}, \nn
\er
\br
{\tiny
a_{21} = \left(\begin{array}{cccccccccc}
        \frac{im^{3}}{2} {\tilde \psi}^{3}_{R} & 
        \frac{im^{3}}{2} {\tilde \psi}^{3}_{R} & 0 & 0 & 0 & 0 & 0 & 0 & 0& 0           
        \\
        \frac{im^{3}}{2} {\tilde \psi}^{3}_{L} & 
        \frac{im^{3}}{2} {\tilde \psi}^{3}_{L} & 0 & 0 & 0 & 0 & 0 & 0 & 0 & 0           
        \\
        0 & 0 & 0 & 0 & 1 & 0 & 0 & 0 & 0 & 0  
        \\
        0 & 0 & 0 & 0 & 0 & 1 & 0 & 0 & 0 & 0 
        \\
        m^1 {\psi}^{1}_{R} & 0 & 0 & 0 & 0 & 0 & -1 & 0 & 0 & 0 
        \\
        m^1 {\psi}^{1}_{L} & 0 & 0 & 0 & 0 & 0 & 0 & -1 & 0 & 0 
        \\
        0 & m^2 {\psi}^{2}_{R} & 0 & 0 & 0 & 0 & 0 & 0 & -1 & 0 
        \\
        0 & m^2 {\psi}^{2}_{L} & 0 & 0 & 0 & 0 & 0 & 0 & 0 & -1
        \\
        \frac{m^{3}}{2} {\psi}^{3}_{R} & 
        \frac{m^{3}}{2} {\psi}^{3}_{R} & 0 & 0 & 0 & 0 & 0 & 0 & 0 & 0 
        \\
        \frac{m^{3}}{2} {\psi}^{3}_{L} & 
        \frac{m^{3}}{2} {\psi}^{3}_{L} & 0 & 0 & 0 & 0 & 0 & 0 & 0 & 0 
        \end{array}\right)}, \,\,
{\tiny
a_{22} = \left(\begin{array}{ccccccc}
        0 & 0 & 0 & \cdots  & 0 & -1 & 0
        \\
        0 & 0 & 0 & \cdots  & 0 & 0 & -1  
        \\
        0 & 0 & 0 & \cdots  & 0 & 0 & 0
        \\
        \multicolumn{7}{c}\dotfill   
        \\
        0 & 0 & 0 &  \cdots  & 0 & 0
& 0
        \\
        -1 & 0 & 0 & \cdots  & 0 & 0 
& 0
        \\
        0 & -1 & 0 & \cdots  & 0 & 0 & 0
        \end{array}\right)}. \nn
\er

The zero-modes are  
\br
        {\bf {{v}}^{(2)}}^T(x) &=& \(u, \; \u, \; \o, \; \chi, \; 0, \; 0, \;
        m^{1} u\psi^{1}_{R}, \; m^{1} u\psi^{1}_{L}, \;
        m^{2} \u\psi^{2}_{R}, \; m^{2} \u\psi^{2}_{L}, \;
        \/{m^{3}}{2}(u+\u)\psi^{3}_{R},                                      \nn
        \\
        &&\/{m^{3}}{2}(u+\u)\psi^{3}_{L}, \; u^{\prime}+\o, \; \u^{\prime}+\chi, \;
        i m^{1}{\ti \psi}^{1}_{R}u, \; im^{1}{\ti \psi}^{1}_{L}u, \;
        im^{2} {\ti \psi}^{2}_{R}\u,                                                  \nn
        \\
        && im^{2} {\ti \psi}^{2}_{L}\u, \; \/{im^{3}}{2}{\ti \psi}^{3}_{R}(u+\u), \;
        i\/{m^{3}}{2}{\ti \psi}^{3}_{L}(u+\u)\).
\er

The zero-mode condition gives no constraints, implying the symmetries of the action
\br
\d {\bf \xi^{(2)}_{A}}\,=\, {\bf v^{(2)}_{A}}(x); \,\,\,\,A=1,2,...20.     
\er

Now, let us choose the gauge conditions 
\be
\lab{gauge2}
\O_{5}\equiv\pi_{1} J_{1}^{1}+\frac{1}{2}J_{1}.(j_{1}+ j_{3})=0, \qq  \O_{6}\equiv \pi_{2} J_{2}^{1}+\frac{1}{2} J_{2}.(j_{2}+ j_{3})=0,
\ee
and impose the consistency conditions with the Lagrange multipliers 
$\eta^{5}, \eta^{6}$, then
\be
{\cal L}^{(3)} = \pi_{1}\dot{\phi}_{1} + \pi_{2}\dot{\phi}_{2} + \dot{\psi}_{R}\pi^{j}_{R}
        + \dot{\psi}_{L}\pi^{j}_{L}  + \dot{\eta}^{1}\O_{1} + \dot{\eta}^{2}\O_{2} 
        + \dot{\eta}^{3}\O_{3} + \dot{\eta}^{4}\O_{4} + \dot{\eta}^{5}\O_{5} 
        + \dot{\eta}^{6}\O_{6}- {\cal V}^{(3)},
\ee
where
\be
{\cal V}^{(3)}={\cal V}^{(2)}|_{\Omega_{5}=\Omega_{6}=0}, 
\ee
or explicitly
\br
{\cal V}^{(3)} &=&  \frac{1}{12}  \( (J_{1})^2 + (J_{2})^2 + J_{1}.J_{2} \)+\frac{1}{2} [ J_{1}.(j_{1}+ j_{3})+ J_{2}.(j_{2}+ j_{3}) ]
   \nn
        \\          
        &+& i \tilde{\psi}^{j}_{R} {\psi}^{j}_{R,x} 
         -  i \tilde{\psi}^{j}_{L} {\psi}^{j}_{L,x}
        + im^{j}_{\psi} \bar{\psi}^{j}\psi^{j}.
\er

The symplectic two-form for this Lagrangian is a non singular matrix, then our 
algorithm has come to an end. Collecting the canonical part and the symplectic potential ${\cal V}^{(3)}$ one has
\be
\lab{thirring2}
{\cal L}[\psi,\overline{\psi}]= \sum_{j=1}^{3}\{i\overline{\psi}^{j}\gamma^{\mu}\pa_{\mu}\psi^{j} 
        + m^{j}_{\psi}\,\,{\overline{\psi}}^{j}\psi^{j}\}\,
        \,-\sum_{{\tiny\begin{array}{c} k,l=1\\k\leq l\end{array}}}^{3}\[ \bar{a}_{kl} j_{k}.j_{l}\] + \sum_{l=1}^{3} m^{l}\nu_{l}\, j_{l}^{0},
\ee
where $\nu_{3} \equiv \frac{\nu_{1}+\nu_{2}}{2}$. We have made the same choice, $k=1/g$, and the field rescalings $\psi^{j} \rightarrow 1/\sqrt{k}\,\psi^{j}$ as in the last section. 
This is the same GMT Lagrangian as \rf{thirring1}. As a bonus, we get the chemical potentials\, $\mu_{l} \equiv m^{l} \nu_{l}$ ( $\dot{\eta}^{1, 2} \rightarrow \nu_{1,2 } $) times the charge densities.
These terms are related to the charges\, $Q_{F}^{l}=\frac{1}{2\pi}\int_{-\infty}^{+\infty} dx\, j_{l}^{0}(t,x)$, and 
their presence is a consequence of the symplectic method \cite{annals}.

\subsection{The generalized sine-Gordon model (GSG)}
\label{sec:sG}

        One can choose other gauge fixings, instead of \rf{gauge1},
        to construct the twice-iterated Lagrangian. Let us make the choice
\be
\lab{gauge3}
\O_{3}\equiv J^{0}_{1} = 0, \qq  \O_{4}\equiv J^{0}_{2} = 0,
\ee
which satisfies the non-gauge invariance condition as can be verified by computing the 
brackets $\{\O_{a} \; , \; J^{0}_{b}\}=0$;\, $a,b=1,2$. The twice-iterated Lagrangian 
is obtained by bringing back these constraints into the canonical part of ${\cal L}^{(1)}$, then
\be
        {\cal L}^{(2)} = \pi_{1}\dot{\phi}_{1} + \pi_{2}\dot{\phi}_{2} + \dot{\psi}_{R}\pi^{j}_{R}
        + \dot{\psi}_{L}\pi^{j}_{L}  + \dot{\eta}^{1}\O_{1} + \dot{\eta}^{2}\O_{2} 
        + \dot{\eta}^{3}\O_{3} + \dot{\eta}^{4}\O_{4} - {\cal V}^{(2)},
\ee
where the twice-iterated symplectic potential becomes
\be
{\cal V}^{(2)}={\cal V}^{(1)}\Big|_{\O_{3}= \O_{4} = 0}. 
\ee

Considering the set of symplectic variables in the following order
\be
\xi^{(2)}_{A} = (\eta^{1}, \eta^{2},\eta^{3}, \eta^{4}, \phi_1, \phi_2, \psi^{1}_{R}, \psi^{1}_{L}, 
        \psi^{2}_{R}, \psi^{2}_{L}, \psi^{3}_{R}, \psi^{3}_{L}, \pi_{1}, \pi_2, \pi^{1}_{R}, 
        \pi^{1}_{L},
\pi^{2}_{R}, \pi^{2}_{L}, \pi^{3}_{R}, \pi^{3}_{L} )
\ee
and the components of the canonical one-form
\be
a^{(2)}_{A} = (\O_1, \O_2, \O_3, \O_4, \pi_{1}, \pi_2, -\pi^{1}_{R}, -\pi^{1}_{L}, -\pi^{2}_{R},
        -\pi^{2}_{L}, -\pi^{3}_{R}, -\pi^{3}_{L}, 0, 0, 0, 0, 0, 0, 0, 0),
\ee
the (degenerated) $20$x$20$ symplectic matrix is found to be
\br
f^{(2)}_{AB}(x,y) &=& \left(\begin{array}{cc}
        a_{11} &  a_{12}                        \\
        a_{21} &  a_{22}                        
        \end{array} \right) \d(x-y),
\er
where
\br     {\tiny
a_{11} = \left(\begin{array}{cccccccccc}
        0 & 0 &0 & 0 &  \pa_{x} & 0 & i m^1 {\tilde \psi}^{1}_{R}  
        & i m^1 {\tilde \psi}^{1}_{L} & 0 & 0              
        \\
        0 & 0 & 0 & 0 & 0 &  \pa_{x} & 0 & 0 & i m^2 {\tilde \psi}^{2}_{R}  &  
        i m^2 {\tilde \psi}^{2}_{L}                
        \\
        0 & 0 & 0 & 0 & 0 & 0 & i m^1 {\tilde \psi}^{1}_{R}  
        & i m^1 {\tilde \psi}^{1}_{L} & 0 & 0  
        \\
        0 & 0 & 0 & 0 & 0 & 0 & 0 & 0 & i m^2 {\tilde \psi}^{2}_{R}  &  
        i m^2 {\tilde \psi}^{2}_{L}        
        \\
         \pa_x & 0 & 0 & 0 & 0 & 0 & 0 & 0 & 0 & 0 
        \\
        0 &  \pa_x & 0 & 0 & 0 & 0 & 0 & 0 & 0 & 0 
        \\
        i m^1 {\tilde \psi}^{1}_{R} & 0 & i m^1 {\tilde \psi}^{1}_{R} 
        & 0 & 0 & 0 & 0 & 0 & 0 & 0
        \\
        i m^1 {\tilde \psi}^{1}_{L} & 0 & i m^1 {\tilde \psi}^{1}_{L}
        & 0 & 0 & 0 & 0 & 0 & 0 & 0
        \\
        0 & i m^2 {\tilde \psi}^{2}_{R} & 0 & i m^2 {\tilde \psi}^{2}_{R}
        &  0 & 0 & 0 & 0 & 0 & 0
        \\
        0 & i m^2 {\tilde \psi}^{2}_{L} & 0 & i m^2 {\tilde \psi}^{2}_{L}
        & 0 & 0 & 0 & 0 & 0 & 0
        \end{array}\right)}, \nn
\er
\br     {\tiny
a_{12} = \left(\begin{array}{cccccccccc}
         \frac{im^{3}}{2} {\tilde \psi}^{3}_{R}  
        & \frac{im^{3}}{2} {\tilde \psi}^{3}_{L} & 0 & 0 & m^1 {\psi}^{1}_{R}  
        & m^1 {\psi}^{1}_{L} & 0 & 0 & \frac{m^{3}}{2} {\psi}^{3}_{R} 
        & \frac{m^{3}}{2} {\psi}^{3}_{L}         
        \\
        \frac{im^{3}}{2} {\tilde \psi}^{3}_{R}
        & \frac{im^{3}}{2} {\tilde \psi}^{3}_{L} & 0 & 0 & 0 & 0 
        & m^2 {\psi}^{2}_{R} & m^2 {\psi}^{2}_{L} 
        &\frac{m^{3}}{2} { \psi}^{3}_{R} &\frac{m^{3}}{2} { \psi}^{3}_{L}                 
        \\
        \frac{im^{3}}{2} {\tilde \psi}^{3}_{R}  
        & \frac{im^{3}}{2} {\tilde \psi}^{3}_{L} & 0 & 0 & m^1 {\psi}^{1}_{R}  
        & m^1 {\psi}^{1}_{L} & 0 & 0 & \frac{m^{3}}{2} {\psi}^{3}_{R} 
        & \frac{m^{3}}{2} {\psi}^{3}_{L} 
        \\
        \frac{im^{3}}{2} {\tilde \psi}^{3}_{R}
        & \frac{im^{3}}{2} {\tilde \psi}^{3}_{L} & 0 & 0 & 0 & 0 
        & m^2 {\psi}^{2}_{R} & m^2 {\psi}^{2}_{L} 
        &\frac{m^{3}}{2} { \psi}^{3}_{R} &\frac{m^{3}}{2} { \psi}^{3}_{L}
        \\
        0 & 0 & -1 & 0 & 0 & 0 & 0 & 0 & 0 & 0 
        \\
        0 & 0 & 0 & -1 & 0 & 0 & 0 & 0 & 0 & 0
        \\
        0 & 0 & 0 & 0 & -1 & 0 & 0 & 0 & 0 & 0 
        \\
        0 & 0 & 0 & 0 & 0 & -1 & 0 & 0 & 0 & 0
        \\
        0 & 0 & 0 & 0 & 0 & 0 & -1 & 0 & 0 & 0   
        \\
        0 & 0 & 0 & 0 & 0 & 0 & 0 & -1 & 0 & 0
        \end{array}\right)}, \nn
\er
\br
{\tiny
a_{21} = \left(\begin{array}{cccccccccc}
        \frac{im^{3}}{2} {\tilde \psi}^{3}_{R} & 
        \frac{im^{3}}{2} {\tilde \psi}^{3}_{R} & 
        \frac{im^{3}}{2} {\tilde \psi}^{3}_{R} & 
        \frac{im^{3}}{2} {\tilde \psi}^{3}_{R} & 0 & 0 & 0 & 0 & 0&0            
        \\
        \frac{im^{3}}{2} {\tilde \psi}^{3}_{L} & 
        \frac{im^{3}}{2} {\tilde \psi}^{3}_{L} & 
        \frac{im^{3}}{2} {\tilde \psi}^{3}_{L} & 
        \frac{im^{3}}{2} {\tilde \psi}^{3}_{L} & 0 & 0 & 0 & 0 & 0 &0           
        \\
        0 & 0 & 0 & 0 & 1 & 0 & 0 & 0 & 0 &0  
        \\
        0 & 0 & 0 & 0 & 0 & 1 & 0 & 0 & 0 &0  
        \\
        m^1 {\psi}^{1}_{R} & 0 & m^1 {\psi}^{1}_{R} & 0 & 0 & 0 & -1 & 0 & 0 &0 
        \\
        m^1 {\psi}^{1}_{L} & 0 & m^1 {\psi}^{1}_{L} & 0 & 0 & 0 & 0 &-1 & 0 &0 
        \\
        0 & m^2 {\psi}^{2}_{R} & 0 & m^2 {\psi}^{2}_{R} & 0 & 0 & 0 & 0 & -1 &0 
        \\
        0 & m^2 {\psi}^{2}_{L} & 0 & m^2 {\psi}^{2}_{L} & 0 & 0 & 0 & 0 & 0 &-1
        \\
        \frac{m^{3}}{2} {\psi}^{3}_{R} & 
        \frac{m^{3}}{2} {\psi}^{3}_{R} & 
        \frac{m^{3}}{2} {\psi}^{3}_{R} & 
        \frac{m^{3}}{2} {\psi}^{3}_{R} & 0 & 0 & 0 & 0 & 0 &0 
        \\
        \frac{m^{3}}{2} {\psi}^{3}_{L} & 
        \frac{m^{3}}{2} {\psi}^{3}_{L} & 
        \frac{m^{3}}{2} {\psi}^{3}_{L} & 
        \frac{m^{3}}{2} {\psi}^{3}_{L} & 0 & 0 & 0 & 0 & 0 &0   
        \end{array}\right)}, \,\,  
{\tiny
a_{22} = \left(\begin{array}{ccccccc}
        0 & 0 & 0 &\cdots & 0 & -1 & 0          
        \\
        0 & 0 & 0 &\cdots & 0 & 0 & -1  
        \\
        0 & 0 & 0 & \cdots & 0 & 0 & 0
        \\
        \multicolumn{7}{c}\dotfill 
        \\
        0 & 0 & 0 &\cdots & 0 & 0 & 0
        \\
        -1 & 0 & 0 & \cdots & 0 & 0 & 0
        \\
        0 & -1 & 0 & \cdots & 0 & 0 & 0
        \end{array}\right)}.                    \nn
\er

Its zero-modes are
\br
{\bf {{v}}^{(2)}}^T(x) &=& \(u, \; \u, \; \o, \; \chi, \; 0, \; 0, \;
        m^{1}(u+\o)\psi^{1}_{R}, \;
        m^{1}(u+\o)\psi^{1}_{L}, \; m^{2}(\u+\chi)\psi^{2}_{R}, \;
        m^{2}(\u+\chi)\psi^{2}_{L},                                          \nn
        \\
        &&\/{m^{3}}{2}(u+\u+\o+\chi)\psi^{3}_{R}, \;
        \/{m^{3}}{2}(u+\u+\o+\chi)\psi^{3}_{L}, \;
        u^{\prime}, \; \u^{\prime}, \;
        im^{1}{\ti \psi}^{1}_{R}(u+\o),                                \nn
        \\
        && im^{1}{\ti \psi}^{1}_{L}(u+\o), \;
        im^{2}{\ti \psi}^{2}_{R}(\u+\chi), \;
        im^{2}{\ti \psi}^{2}_{L}(\u+\chi), \;
        \/{im^{3}}{2}{\ti \psi}^{3}_{R}(u+\u+\o+\chi),                           \nn
        \\
        && \/{im^{3}}{2}{\ti \psi}^{3}_{L}(u+\u+\o+\chi)\),
\er
where $u$, $\u$, $\o$ and $\chi$ are arbitrary functions. The zero mode condition becomes
\br
\int {dx {{\bf v}^{(2)}}^{T}(x) \frac{\d}{\d \xi^{(2)}}\int{dy'{\cal V}^{(2)}}} 
        = \int{dx\, J^{1}_{a}\, \partial_{x} f_{a}} \equiv 0,\,\,\,\,\,\,f_{a}\equiv (\o, \chi),\,\,\,a=1,2.\nn
\er

Since the functions $f_{a}$ are arbitrary we end up with the following constraints
\be
\lab{lagcons}
        \O_{5} \equiv J^{1}_1=0,\,\,\,\,\,\, \O_{6} \equiv J^{1}_2 = 0.
\ee 

Notice that by solving the constraints, $\O_{3}=\O_{4}=\O_{5}=\O_{6}=0$, Eqs. \rf{gauge3} and \rf{lagcons}, we may obtain
\be
\lab{majorana}
        {\ti \psi}^{j}_{R} = \psi^{j}_{R}, \qq {\ti \psi}^{j}_{L} = \psi^{j}_{L}.
\ee 

So, at this stage, we have Majorana spinors, the scalars $\phi_{1}$ and $\phi_{2}$, and the auxiliary fields. 
Next, introduce a third set of Lagrange multipliers into
${\cal L}^{(2)}$, then
\be
{\cal L}^{(3)} = \pi_{1}\dot{\phi}_{1} + \pi_{2}\dot{\phi}_{2} + \dot{\psi}_{R}\pi^{j}_{R}
        + \dot{\psi}_{L}\pi^{j}_{L}  + \dot{\eta}^{1}\O_{1} + \dot{\eta}^{2}\O_{2} 
        + \dot{\eta}^{3}\O_{3} + \dot{\eta}^{4}\O_{4} + \dot{\eta}^{5}\O_{5} 
        + \dot{\eta}^{6}\O_{6}- {\cal V}^{(3)},
\ee
where
\be
{\cal V}^{(3)}={\cal V}^{(2)}|_{\Omega_{5} = \Omega_{6} = 0}
\ee
or
\be
        {\cal V}^{(3)} = {1\over 24}{\phi_{j,x}}^2 + i{\psi}^{j}_{R}\psi^{j}_{R,x}
                    - i{\psi}^{j}_{L}\psi^{j}_{L,x} 
                    + i m^{j}_{\psi}{\psi}^{j}_{R} \psi^{j}_{L} (e^{-i\phi_{j}}+ e^{i\phi_{j}}).
\ee

The new set of symplectic variables is assumed to be ordered as
\br
        \xi^{(3)}_{A} = (\eta^{1}, \eta^{2},\eta^{3}, \eta^{4}, \eta^{5}, \eta^{6}, \phi_1, 
        \phi_2, \psi^{1}_{R}, \psi^{1}_{L}, \psi^{2}_{R}, \psi^{2}_{L}, \psi^{3}_{R},
        \psi^{3}_{L}, \pi_{1}, \pi_2, \pi^{1}_{R}, \pi^{1}_{L},
\pi^{2}_{R}, \pi^{2}_{L},
        \pi^{3}_{R}, \pi^{3}_{L} ).\nn
\er

The components of the canonical one-form are
\br
        a_{A}^{(3)} = (\O_1, \O_2, \O_3, \O_4, \O_5, \O_6, \pi_{1}, \pi_2, -\pi^{1}_{R}, 
        -\pi^{1}_{L}, -\pi^{2}_{R}, -\pi^{2}_{L}, -\pi^{3}_{R}, -\pi^{3}_{L}, 
        0, 0, 0, 0, 0, 0, 0, 0).\nn 
\er

After some algebraic manipulations we get the third-iterated $22$x$22$ symplectic two-form
\br
\lab{forma}
        f^{(3)}_{AB}(x,y) &=& \left(\begin{array}{cc}
        a_{11} &  a_{12}                        \\
        a_{21} &  a_{22}                        
        \end{array} \right) \d(x-y).
\er
where 
\br
        {\tiny a_{11} = \left(\begin{array}{ccccccccccc}
        0 & 0 & 0 & 0 & 0 & 0 &  \pa_{x} & 0 & i m^1 {\tilde \psi}^{1}_{R}  
        & i m^1 {\tilde \psi}^{1}_{L} & 0          
        \\
        0 & 0 & 0 & 0 & 0 & 0 & 0 &  \pa_{x} & 0 & 0 & i m^2 {\tilde \psi}^{2}_{R}              
        \\
        0 & 0 & 0 & 0 & 0 & 0 & 0 & 0 & i m^1 {\tilde \psi}^{1}_{R}  
        & i m^1 {\tilde \psi}^{1}_{L} & 0   
        \\
        0 & 0 & 0 & 0 & 0 & 0 & 0 & 0 & 0 & 0 & i m^2 {\tilde \psi}^{2}_{R}  
        \\
        0 & 0 & 0 & 0 & 0 & 0 & 0 & 0 & -i m^1 {\tilde \psi}^{1}_{R}
        & i m^1 {\tilde \psi}^{1}_{L} & 0
        \\
        0 & 0 & 0 & 0 & 0 & 0 & 0 & 0 & 0 & 0 & -i m^2 {\tilde \psi}^{2}_{R}
        \\
         \pa_x & 0 & 0 & 0 & 0 & 0 & 0 & 0 & 0 & 0 & 0
        \\
        0 &  \pa_x & 0 & 0 & 0 & 0 & 0 & 0 & 0 & 0 & 0
        \\
        i m^1 {\tilde \psi}^{1}_{R} & 0 & i m^1 {\tilde \psi}^{1}_{R} 
        & 0 & -i m^1 {\tilde \psi}^{1}_{R} & 0 & 0 & 0 & 0 & 0 & 0
        \\
        i m^1 {\tilde \psi}^{1}_{L} & 0 & i m^1 {\tilde \psi}^{1}_{L}
        & 0 & i m^1 {\tilde \psi}^{1}_{L} & 0 & 0 & 0 & 0 & 0 & 0
        \\
        0 & i m^2 {\tilde \psi}^{2}_{R} & 0 & i m^2 {\tilde \psi}^{2}_{R}
        & 0 & -i m^2 {\tilde \psi}^{2}_{R} & 0 & 0 & 0 & 0 & 0
        \end{array}\right)},                                                    \nn
\er

\br
        {\tiny a_{12}=\left(\begin{array}{ccccccccccc}
        0 & \frac{im^{3}}{2} {\tilde \psi}^{3}_{R}  
        & \frac{im^{3}}{2} {\tilde \psi}^{3}_{L} & 0 & 0 & m^1 {\psi}^{1}_{R}  
        & m^1 {\psi}^{1}_{L} & 0 & 0 & \frac{m^{3}}{2} {\psi}^{3}_{R} 
        & \frac{m^{3}}{2} {\psi}^{3}_{L}         
        \\
        i m^2 {\tilde \psi}^{2}_{L} & \frac{im^{3}}{2} {\tilde \psi}^{3}_{R}
        & \frac{im^{3}}{2} {\tilde \psi}^{3}_{L} & 0 & 0 & 0 & 0 
        & m^2 {\psi}^{2}_{R} & m^2 {\psi}^{2}_{L} 
        &\frac{m^{3}}{2} { \psi}^{3}_{R} &\frac{m^{3}}{2} { \psi}^{3}_{L}                 
        \\
        0 & \frac{im^{3}}{2} {\tilde \psi}^{3}_{R}  
        & \frac{im^{3}}{2} {\tilde \psi}^{3}_{L} & 0 & 0 & m^1 {\psi}^{1}_{R}  
        & m^1 {\psi}^{1}_{L} & 0 & 0 & \frac{m^{3}}{2} {\psi}^{3}_{R} 
        & \frac{m^{3}}{2} {\psi}^{3}_{L} 
        \\
        i m^2 {\tilde \psi}^{2}_{L} & \frac{im^{3}}{2} {\tilde \psi}^{3}_{R}
        & \frac{im^{3}}{2} {\tilde \psi}^{3}_{L} & 0 & 0 & 0 & 0 
        & m^2 {\psi}^{2}_{R} & m^2 {\psi}^{2}_{L} 
        &\frac{m^{3}}{2} { \psi}^{3}_{R} &\frac{m^{3}}{2} { \psi}^{3}_{L}
        \\
        0 & -\frac{im^{3}}{2} {\tilde \psi}^{3}_{R} 
           & \frac{im^{3}}{2} {\tilde \psi}^{3}_{L} & 0 & 0 
           & -m^1 {\psi}^{1}_{R} & m^1 {\psi}^{1}_{L} & 0 & 0 
           & -\frac{m^{2}}{2} \psi^{3}_{R} 
           &  \frac{m^{3}}{2} \psi^{3}_{L}  
        \\
         i m^2 {\tilde \psi}^{2}_{L} & -\frac{im^{3}}{2} {\tilde \psi}^{3}_{R} 
           & \frac{im^{3}}{2} {\tilde \psi}^{3}_{L} & 0 & 0 & 0 & 0 
           & -m^2 {\psi}^{2}_{R} & m^2 {\psi}^{2}_{L} 
           & -\frac{m^{3}}{2} \psi^{3}_{R} 
           &  \frac{m^{3}}{2} \psi^{3}_{L}  
        \\
        0 & 0 & 0 & -1 & 0 & 0 & 0 & 0 & 0 & 0 & 0 
        \\
        0 & 0 & 0 & 0 & -1 & 0 & 0 & 0 & 0 & 0 & 0 
        \\
        0 & 0 & 0 & 0 & 0 & -1 & 0 & 0 & 0 & 0 & 0 
        \\
        0 & 0 & 0 & 0 & 0 & 0 & -1 & 0 & 0 & 0 & 0 
        \\
        0 & 0 & 0 & 0 & 0 & 0 & 0 & -1 & 0 & 0 & 0 
        \end{array}\right)},                            \nn
\er
\br
        {\tiny a_{21} = \left(\begin{array}{ccccccccccc}
        0 & i m^2{\tilde \psi}^{2}_{L} &
        0 & i m^2{\tilde \psi}^{2}_{L} & 0 & i m^2 {\tilde \psi}^{2}_{L} & 
        0 & 0 & 0 & 0 & 0 
        \\
        \frac{im^{3}}{2} {\tilde \psi}^{3}_{R} & 
        \frac{im^{3}}{2} {\tilde \psi}^{3}_{R} & 
        \frac{im^{3}}{2} {\tilde \psi}^{3}_{R} & 
        \frac{im^{3}}{2} {\tilde \psi}^{3}_{R} & 
       -\frac{im^{3}}{2} {\tilde \psi}^{3}_{R} & 
       -\frac{im^{3}}{2} {\tilde \psi}^{3}_{R} &
        0 & 0 & 0 & 0 & 0               
        \\
        \frac{im^{3}}{2} {\tilde \psi}^{3}_{L} & 
        \frac{im^{3}}{2} {\tilde \psi}^{3}_{L} & 
        \frac{im^{3}}{2} {\tilde \psi}^{3}_{L} & 
        \frac{im^{3}}{2} {\tilde \psi}^{3}_{L} &
        \frac{im^{3}}{2} {\tilde \psi}^{3}_{L} & 
        \frac{im^{3}}{2} {\tilde \psi}^{3}_{L} & 
        0 & 0 & 0 & 0 & 0               
        \\
        0 & 0 & 0 & 0 & 0 & 0 & 1 & 0 & 0 & 0 & 0 
        \\
        0 & 0 & 0 & 0 & 0 & 0 & 0 & 1 & 0 & 0 & 0
        \\
        m^1 {\psi}^{1}_{R} & 0 & 
        m^1 {\psi}^{1}_{R} & 0 & 
       -m^1 {\psi}^{1}_{R} & 0 & 0 & 0 & -1 & 0 & 0
        \\
        m^1 {\psi}^{1}_{L} & 0 & 
        m^1 {\psi}^{1}_{L} & 0 &  
        m^1 {\psi}^{1}_{L} & 0 & 0 & 0 & 0 & -1 & 0 
        \\
        0 & m^2 {\psi}^{2}_{R} & 0 & 
            m^2 {\psi}^{2}_{R} & 0 & 
            -m^2 {\psi}^{2}_{R} & 0 & 0 & 0 & 0 & -1
        \\
        0 & m^2 {\psi}^{2}_{L} & 0 & 
            m^2 {\psi}^{2}_{L} & 0 & 
           m^2 {\psi}^{2}_{L} & 0 & 0 & 0 & 0 & 0
        \\
        \frac{m^{3}}{2} {\psi}^{3}_{R} & 
        \frac{m^{3}}{2} {\psi}^{3}_{R} & 
        \frac{m^{3}}{2} {\psi}^{3}_{R} & 
        \frac{m^{3}}{2} {\psi}^{3}_{R} & 
       -\frac{m^{3}}{2} {\psi}^{3}_{R} & 
       -\frac{m^{3}}{2} {\psi}^{3}_{R} & 0 & 0 & 0 & 0 & 0 
        \\
        \frac{m^{3}}{2} {\psi}^{3}_{L} & 
        \frac{m^{3}}{2} {\psi}^{3}_{L} & 
        \frac{m^{3}}{2} {\psi}^{3}_{L} & 
        \frac{m^{3}}{2} {\psi}^{3}_{L} & 
        \frac{m^{3}}{2} {\psi}^{3}_{L} & 
        \frac{m^{3}}{2} {\psi}^{3}_{L} & 0 & 0 & 0 & 0 & 0
        \end{array}\right)},                                            \nn
\er
\br
        {\tiny a_{22} = \left(\begin{array}{ccccccccc}
        0 & 0 & 0 & 0 &\cdots & 0 & -1 & 0 & 0              
        \\
        0 & 0 & 0 & 0 &\cdots & 0 & 0 & -1 & 0
        \\
        0 & 0 & 0 & 0 & \cdots & 0 & 0 & 0 & -1
        \\
        0 & 0 & 0 & 0 &\cdots & 0 & 0 & 0 & 0
        \\
        \multicolumn{9}{c}\dotfill
        \\
        0 & 0 & 0 & 0 &\cdots & 0 & 0 & 0 & 0
        \\
        -1 & 0 & 0 & 0 &\cdots & 0 & 0 & 0 & 0
        \\
        0 & -1 & 0 & 0 &\cdots & 0 & 0 & 0 & 0
        \\
        0 & 0 & -1 & 0 &\cdots & 0 & 0 & 0 & 0 
        \end{array}\right)}.                    \nn
\er

It can be checked that this matrix has the zero-modes
\br
        {\bf {v}}^{(3)}(x) &=& (u, \; \u, \; \o, \; \chi, \; y, \; z, \; 0, \; 0, \;
        m^1 a^-_1\psi^1_R, \;
        m^1 a^+_1\psi^1_L, \;
        m^2 a^-_2\psi^2_R, \;
        m^2 a^+_2\psi^2_L, \;
        \frac{m^3}{2} a^{-}_3\psi^3_R, 
                \nn
        \\
        && \frac{m^3}{2} a^{+}_3 \psi^3_L, \;  u^{\prime}, \;  \u^{\prime}, \;
        i m^1 a^-_1{\ti \psi}^1_R, \;
        i m^1 a^+_1{\ti \psi}^1_L, \;
        i m^2 a^-_2{\ti \psi}^2_R, \;
        i m^2 a^+_2{\ti \psi}^2_L, \;                      \nn
        \\
        && i\frac{m^3}{2} a^{-}_3{\ti \psi}^3_R, \;
        i\frac{m^3}{2} a^+_3{\ti \psi}^3_L ), 
\er
where $a^+_1 \equiv u+\o+y,\, a^+_2 \equiv \u+\chi+z,\, a^+_3 \equiv  u+\o+y+\u+\chi+z,\, a^-_1 \equiv u+\o-y, \, a^-_2 \equiv \u+\chi-z, \, a^+_3 \equiv  u+\o+y+\u-\chi-z$, and $u$, $\u$, $\o$, $\chi$, $y$\, and $z$ are arbitrary functions. The relevant zero-mode condition gives no constraints. Then the action has the following symmetries
\br
\lab{symmetry1}
\d \xi^{(3)}_{A}\,=\, {\bf v^{(3)}_{A}}(x); \,\,\,\,A=1,2,...22. 
\er

These symmetries allow us to fix the bilinears $i\psi^j_{R}\psi^j_{L}$ to be constants. By taking 
$\psi_{R}^{j}=-iC_{j} \overline{\theta}_{j}$ and $\psi_{L}^{j}=\theta_{j}$, $(j=1,2,3)$ with $C_{j}$ being real numbers, we find that
$i\psi_{R}^{j}\psi_{L}^{j}$ indeed becomes a constant. Note that $\theta_{j}$ and $\overline{\theta}_{j}$ are 
Grassmannian variables, while $\overline{\theta}_{j}\theta_{j}$ is an ordinary commuting number. 

The two form $f_{AB}^{(3)}(x,y)$, Eq. \rf{forma}, in the subspace 
$(\phi_{1}, \phi_{2}, \pi_{\phi_{1}}, \pi_{\phi_{2}})$ defines a canonical symplectic structure modulo canonical transformations.
The coordinates $\phi_{a}$ and $\pi_{\phi_{a}}$\, ($a=1,2$) are not unique. Consider a canonical transformation from
$(\phi_{a}, \pi_{\phi_{a}})$ to $(\hat{\phi_{a}}, \hat{\pi}_{\hat{\phi_{a}}})$ such that
$\phi_{a}=\frac{\pa F}{\pa \pi_{\phi_{a}}}$ and $\hat{\phi_{a}}=\frac{\pa F}{\pa \hat{\pi}_{\hat{\phi}_{a}} }$. Then, in particular if $\phi_{a}=\hat{\phi_{a}}$ one can, in principle, solve for the 
function $F$ such that a manifestly kinetic termr appear in the new Lagrangian. 
    
Then choosing $k=1/g$ as the overall coupling constant, we are left with
\be
\lab{sine}
{\cal L^{''}}=\sum_{j=1}^{3}\[\frac{1}{24 g} \partial_{\mu}\phi_{j} \partial^{\mu}\phi_{j}
        + \frac{M_{j}}{g} \; \mbox{cos} \phi_{j}\]+ \mu_{1}\pa_{x} \phi_{1}+\mu_{2} \pa_{x}\phi_{2},
\ee
where $M_{j}=m_{\psi}^{j}C_{j}$. This defines the {\sl  generalized sine-Gordon model} (GSG). In addition we get the terms multiplied by chemical potentials $\mu_{1}$ and $\mu_{2}$ ($\dot{\eta}^{1,\,2} \rightarrow -\mu_{1,\,2}$). These are just the topological charge densities, and are related to the conservation of the number of kinks minus antikinks $Q_{\rm topol.}^{a}= \frac{1}{\pi} \int_{-\infty}^{+\infty} dx\,\,\pa_{x} \phi_{a}$.  
  
In the above gauge fixing procedures the possibility of Gribov-like ambiguities deserves a careful analysis. See Ref. \cite{annals} for a discussion in the $sl(2)$ ATM case. However, in the next section, we provide an indirect evidence of the abscence of such ambiguities, at least, for the soliton sector of the model.

\section{The soliton/particle correspondences}
\label{sec:dual}

The $sl(2)$ ATM theory contains the sine-Gordon (SG) and the massive Thirring (MT) models
describing the soliton/particle correspondence of its spectrum \cite{bla,annals,witten}. The ATM one-(anti)soliton solution
satisfies the remarkable SG and MT classical correspondence in which, apart from the Noether and topological currents equivalence, MT spinor bilinears are related to the exponential of the SG field \cite{orfanidis}. The last relationship was exploited in \cite{nucl} to decouple the $sl(2)$ ATM equations of
motion into the SG and MT ones. Here we provide a generalization of that
correspondence to the $sl(3)$ ATM case. In fact, consider the relationships 
\begin{eqnarray}
\nn
\frac{\psi _{R}^{1}\widetilde{\psi}_{L}^{1}}{i} &=&-\frac{1}{4\D}[\left(
m^{1}_{\psi} p_{1}-m^{3}_{\psi}p_{4}-m^{2}_{\psi}p_{5}\right) e^{i\left( \varphi _{2}-2\varphi_{1}\right) }+m^{2}_{\psi}p_{5}e^{3i\left( \varphi _{2}-\varphi _{1}\right) }  \\
&&+m^{3}_{\psi}p_{4}e^{-3i\varphi _{1}}-m^{1}_{\psi}p_{1}] \lab{duality31}\\
\nn
\psi _{R}^{2}\widetilde{\psi}_{L}^{2} &=&-\frac{1}{4\D}[\left(
m^{2}_{\psi}p_{2}-m^{1}_{\psi}p_{5}-m^{3}_{\psi}p_{6}\right) e^{i\left( \varphi _{1}-2\varphi_{2}\right) }+m^{1}_{\psi}p_{5}e^{3i\left( \varphi _{1}-\varphi _{2}\right) } 
\\
&&-m^{3}_{\psi}p_{6}e^{-3i\varphi _{2}}-m^{2}_{\psi}p_{2}]
\lab{duality32} \\
\nn
\frac{\widetilde{\psi}_{R}^{3}\psi_{L}^{3}}{i} &=&-\frac{1}{4\D}[\left(
m^{3}_{\psi}p_{3}-m^{1}_{\psi}p_{4}+m^{2}_{\psi}p_{6}\right) e^{i\left( \varphi_{1}+\varphi_{2}\right) }+m^{1}_{\psi}p_{4}e^{3i\varphi _{1}}  \\
&&-m^{2}_{\psi}p_{6}e^{3i\varphi _{2}}-m^{3}_{\psi}p_{3}],  \lab{duality33}
\end{eqnarray}
where $\D \equiv \bar{a}_{11}\bar{a}_{22}\bar{a}_{33}+2\bar{a}_{12}\bar{a}_{23}\bar{a}_{13}-\bar{a}_{11}\left(
\bar{a}_{23}\right)^{2}-\left( \bar{a}_{12}\right)^{2}\bar{a}_{33}-\left( \bar{a}_{13}\right)
^{2}\bar{a}_{22}$;\,  $p_{1}\equiv \left( \bar{a}_{23}\right)^{2}-\bar{a}_{22}\bar{a}_{33}$;\, $p_{2}\equiv \left( \bar{a}_{13}\right) ^{2}-\bar{a}_{11}\bar{a}_{33}$;\, $p_{3}\equiv \left(
\bar{a}_{12}\right)^{2}-\bar{a}_{11}\bar{a}_{22}$;\, $p_{4}\equiv \bar{a}_{12}\bar{a}_{23}-\bar{a}_{22}\bar{a}_{13}$;\, $p_{5}\equiv \bar{a}_{13}\bar{a}_{23}-\bar{a}_{12}\bar{a}_{33}$;\, $p_{6}\equiv \bar{a}_{11}\bar{a}_{23}-\bar{a}_{12}\bar{a}_{13}$ and the $\bar{a}_{ij}$'s being the current-current coupling constants of the GMT model \rf{thirring1}. The relationships \rf{duality31}-\rf{duality33} supplied with the conditions \rf{real1}-\rf{condi} and conveniently substituted into Eqs. \rf{eqnm1}
and \rf{eqnm4}-\rf{eqnm15} decouple the $sl(3)^{(1)}$ CATM equations into the GSG \rf{sine} and GMT \rf{thirring1} equations of motion, respectively.

Moreover, one can show that the GSG \rf{sine} $M_{j}$ parameters and the GMT \rf{thirring1} couplings $\bar{a}_{ij}$ are related by
\begin{eqnarray}
\frac{2\D M_{1}}{g (m^{1}_{\psi})^{2}} &=&\bar{a}_{22}(-\frac{m^{3}_{\psi}}{m^{1}_{\psi}}
\bar{a}_{13}+\bar{a}_{33})+\bar{a}_{23}(-\bar{a}_{23}+\frac{m^{3}_{\psi}}{m^{1}_{\psi}}\bar{a}_{12}), \lab{strongweak31} \\
\frac{2\D M_{2}}{g (m^{2}_{\psi})^{2}} &=&\bar{a}_{11}(-\frac{m^{3}_{\psi}}{m^{2}_{\psi}}
\bar{a}_{23}+\bar{a}_{33})+\bar{a}_{13}(-\bar{a}_{13}+\frac{m^{3}_{\psi}}{m^{2}_{\psi}}\bar{a}_{12}),  \lab{strongweak32} \\
\frac{2\D M_{3}}{g (m^{3}_{\psi})^{2}} &=&-\frac{m^{1}_{\psi}m^{2}_{\psi}}{(m^{3}_{\psi})^{2}}
(\bar{a}_{12}\bar{a}_{33}-\bar{a}_{13}\bar{a}_{23})-\bar{a}_{11}\bar{a}_{22}+(\bar{a}_{12})^{2}. \lab{strongweak33}
\end{eqnarray}

Various limiting cases of the relationships \rf{duality31}-\rf{duality33} and \rf{strongweak31}-\rf{strongweak33} are possible. First, let us consider 
\br
\lab{limit1}
\bar{a}_{jk} \rightarrow \left\{\begin{array}{ll}\infty & \,\,\,j=k \neq l,\,\,\, (\mbox{for a given}\,\,\, l) \\ 
 \mbox{finite} & \,\,\,\mbox{other cases}
\end{array}
\right\}
\er
then one has
\br
\frac{\psi _{R}^{l}\widetilde{\psi}_{L}^{l}}{i}&=&\frac{m^{l}_{\psi}}{4}\left( e^{-i\phi_{l}}-1\right);\,\,\,\,\, \psi _{R}^{j}\widetilde{\psi}_{L}^{j}=0,\,\,\,\,\,\,\,\,j\neq l,
\lab{duality}
\er
for  $\bar{a}_{ll}=\d_{l} g \, (\d_{1,2}=1,\,\, \d_{3}=-1)$. The three species of one-soliton solutions of the $sl(3)$ ATM theory \rf{atm1}, found in \cite{bueno} and described in Section \ref{sec:ATM}, satisfy the relationships \rf{duality} \cite{nucl}. Moreover,
from Eqs. \rf{strongweak31}-\rf{strongweak33} taking the same limits as
in \rf{limit1} one has
\begin{equation}
M_{l}=\frac{(m^{l}_{\psi})^{2}}{2}; \,\,\,\, M_{j}=0,\,\,\,j \neq l.  \lab{weakstrong}
\end{equation}

Therefore, the relationships \rf{duality31}-\rf{duality33} incorporate each $sl(2)$ ATM submodel (particle/soliton) 
weak/strong coupling phases; i.e., the MT/SG correspondence \cite{nucl,annals}.

Then, the currents equivalence \rf{equivalence}, the relationships \rf{duality31}-\rf{duality33} and the conditions \rf{real1}-\rf{condi} satisfied by the one-soliton sector of CATM theory allowed us to stablish the correspondence between the GSG and GMT models, thus extending the MT/SG result \cite{orfanidis}. It could be interesting to obtain the counterpart of Eqs. \rf{duality31}-\rf{duality33} for the $N_{S} \geq 2$ solitons, for example along the lines of \cite{orfanidis}. For $N_{S}=2$, Eq. \rf{equivalence} still holds \cite{bueno}; and Eqs. \rf{real1}-\rf{condi} are satisfied for the species $(\a_{i},\a_{i})$.  

Second, consider the limit
\br
\lab{limit2}
\bar{a}_{ik} \rightarrow \left\{\begin{array}{ll} \infty & \,\,\,i=k=j, \,\,\,\,(\mbox{for a chosen}\, j;\,\,\,j=1,2)\,\,\, 
\\ 
 \mbox{finite} & \,\,\,\mbox{other cases}
\end{array}
\right\}
\er
one gets $M_{j}\,=\,0$ and 
\br
4 \bar{\D}\frac{\psi _{R}^{l}\widetilde{\psi}_{L}^{l}}{i}&=& (m_{\psi}^{3}\bar{a}_{l3}-m_{\psi}^{l}\bar{a}_{33}) e^{-i\phi_{l}}-m_{\psi}^{3} \bar{a}_{l3} e^{-3i\varphi_{l}}+ m_{\psi}^{l} \bar{a}_{33},\,\,\, l\neq j \nn\\
4 \bar{\D}\frac{\psi _{R}^{3}\widetilde{\psi}_{L}^{3}}{i}&=&(m_{\psi}^{3}\bar{a}_{ll}-m_{\psi}^{l}\bar{a}_{l3})  e^{-i\phi_{3}}+m_{\psi}^{l} \bar{a}_{l3} e^{-3i\varphi_{l}}+ m_{\psi}^{3} \bar{a}_{ll},\\
\psi _{R}^{j}\widetilde{\psi}_{L}^{j}&=&0.
\lab{duality2}
\er
where $\bar{\D}\equiv 4(\bar{a}_{ll}\bar{a}_{33}-(\bar{a}_{l3})^2)$.  The parameters are related by $ (m_{\psi}^{3})^2 \bar{a}_{ll} M_{l} = m_{\psi}^{l}(m_{\psi}^{3}\bar{a}_{l3}-m_{\psi}^{l}\bar{a}_{33})  M_{3}$. In the case $M_{l}=M_{3}=M$ and redefining the fields as $\phi_{l}=\sqrt{12 g}(A+B),\, \phi_{j}=-\sqrt{12 g} B$ in the GSG sector, one gets the Lagrangian
\br
 {\cal L}_{BL}= \frac{1}{2}(\pa_{\mu} A)^2+ \frac{1}{2}(\pa_{\mu} B)^2+ 2\frac{M}{g} \mbox{cos}\sqrt{24 g} A\,\, \mbox{cos}\sqrt{72 g} B,
\er
which is a particular case of the Bukhvostov-Lipatov model (BL) \cite{bukhvostov}. It corresponds to a GMT-like theory with two Dirac spinors. The BL model is not classically integrable \cite{ameduri}, and some discussions have appeared in the literature about its quantum integrability \cite{saleur}. 

Alternatively, if one allows the limit $\bar{a}_{33} \rightarrow \infty$ one gets $\psi _{R}^{3}\widetilde{\psi}_{L}^{3}\,=\,0$, and additional relations for the $\psi^{1}, \psi^{2}$ spinors and the $\varphi_{a}$ scalars. The parameters are related by $\frac{M_{1}}{(m_{\psi}^{1})^2 \bar{a}_{22}}=\frac{M_{2}}{(m_{\psi}^{2})^2 \bar{a}_{11}}=-\frac{M_{3}}{m_{\psi}^{1} m_{\psi}^{2} \bar{a}_{12}}$. Then we left with two Dirac spinors in the GMT sector and all the terms of the GSG model. The later resembles the $2-$cosine model studied in \cite{gerganov} in some submanifold of its renormalized parameter space.

\section{Generalization to higher rank Lie algebra}
\label{sec:sl(n)}

The procedures presented so far can directly be extended to the CATM model for the affine Lie algebra $sl(n)^{(1)}$ furnished with the principal gradation. According to the construction of \cite{matter}, these models have soliton solutions for an off-critical submodel, possess a $U(1)$ vector current proportional to a topological current, apart from the conformal symmetry they exhibit a $\(U(1)_{R}\)^{n-1}\otimes \(U(1)_{L}\)^{n-1}$ left-right local gauge symmetry, and the equations of motion describe the dynamics of the scalar fields $\vp_{a},\, \eta,\, \widetilde{\nu}\, (a=1,...n-1)$ and the Dirac spinors $\psi^{\a_{j}}$, $\widetilde{\psi}^{\a_{j}}$, ($j=1,...N$; $N\equiv \frac{n}{2}(n-1)$ = number of positive roots\, $\a_{j}$\, of the simple Lie algebra $sl(n)$) with one-(anti)soliton solution associated to the field\, $\a_{j}.\vec{\vp}$\, ($\vec{\vp}=\sum_{a=1}^{n-1}\vp_{a} \a_{a}$,\, $\a_{a}$= simple roots of $sl(n)$)  for each pair of Dirac fields ($\psi^{\a_{j}}$, $\widetilde{\psi}^{\a_{j}}$)\cite{matter}. Therefore, it is possible to define the off-critical real Lagrangian $sl(n)$ ATM model for  the solitonic sector of the theory. The reality conditions would generalize the Eqs. \rf{real1}-\rf{condi}; i.e., the new $\vp$'s real and the identifications $\widetilde{\psi}^{\a_{j}} \sim (\psi^{\a_{j}})^{*}$ (up to $\pm$ signs). To apply the symplectic analysis of $sl(n)$ ATM one must impose ($n-1$) constraints in the Lagrangian, analogous to \rf{lagrangian1}, due to the above local symmetries. The outcome will be a parent Lagrangian of a generalized massive Thirring model (GMT) with $N$ Dirac fields and a generalized sine-Gordon model (GSG) with ($n-1$) fields. The decoupling of the Toda fields and Dirac fields in the equations of motion of $sl(n)^{(1)}$ CATM, analogous to \rf{eqnm1}
and \rf{eqnm4}-\rf{eqnm15}, could be performed by an extension of the relationships \rf{duality31}-\rf{duality33} and \rf{real1}-\rf{condi}.

\section{Discussions and outlook}
\label{sec:outlook}

We have shown, in the context of FJ and symplectic methods, that the $sl(3)$ ATM \rf{atm1} theory is a parent Lagrangian \cite{hje} from which both the GMT \rf{thirring1} and the GSG \rf{sine} models are derivable. From \rf{thirring1} and \rf{sine}, it is also clear the duality exchange of the couplings: $g \rightarrow 1/g$. The various soliton/particle species correspondences are uncovered. The soliton sector satifies the $U(1)$ vector and topological currents equivalence \rf{equivalence} and decouples the equations of motion into both dual sectors, through the relationships \rf{duality31}-\rf{duality33} (supplied with \rf{real1}-\rf{condi}). The relationships \rf{duality31}-\rf{duality33} contain each $sl(2)$ ATM submodel soliton solution. In connection to these points, recently a parent Lagrangian method was used to give a generalization of the dual theories concept for non $p$-form fields \cite{casini}. In \cite{casini}, the parent Lagrangian contained both types of fields, from which each dual theory was obtained by eliminating the other fields through the equations of motion.

On the other hand, in nonabelian bosonization of massless fermions \cite{witten1}, the fermion bilinears are identified with bosonic operators. Whereas, in abelian bosonization \cite{coleman} there exists an identification between the massive fermion operator (charge nonzero sector) and a nonperturbative bosonic soliton operator \cite{mandelstam}. Recently, it has been shown that symmetric space sine-Gordon models bosonize the massive nonabelian (free) fermions providing the relationships between the fermions and the relevant solitons of the bosonic model \cite{park}. The ATM model allowed us to stablish these type of relationships for interacting massive spinors in the spirit of particle/soliton correspondence. We hope that the quantization of the ATM theories and the related WZNW models, and  in particular the relationships \rf{relationships}, would provide the bosonization of the nonzero charge sectors of the GMT fermions in terms of their associated Toda and WZNW fields. In addition, the above approach to the GMT/GSG duality may be useful to construct the conserved currents and the algebra of their associated charges in the context of the CATM $\rightarrow$ ATM reduction. These currents in the MT/SG case were constructed treating each model as a perturbation on a conformal field theory (see \cite{kaul} and references therein).

Moreover, two dimensional models with four-fermion interactions have played an important role in the understanding of QCD (see, e.g. \cite{bennett} and references therein). Besides, the GMT model contains explicit mass terms: most integrable models such as the Gross-Neveu, $SU(2)$ and $U(1)$ Thirring  models rather all present spontaneous mass generation, the exception being the massive Thirring model. A GMT submodel with $a_{ii}=0$, $a_{ij}=1\, (i>j)$ and equal $m_{\psi}^{j}$'s, defines the so-called extended Bukhvostov-Lipatov model (BL) and has recently been studied by means of a bosonization technique \cite{sakamoto1}. Finally, BL type models were applied to $N-$body problems in nuclear physics \cite{sakamoto}.

\vspace{1cm}
 
\noindent {\bf Acknowledgements}

HB thanks Professors L.A. Ferreira and A.H. Zimerman for discussions on integrable models and Professors M.B. Halpern and R.K. Kaul for correspondences and valuable comments on the manuscript. The authors thank Professors B.M. Pimentel for discussions on constrained systems and A.J. Accioly for encouragement. The authors are supported by FAPESP.

\appendix

\section{The $sl(3)^{(1)}$ CATM model }
\label{app:catm}
We summarize the construction and some properties of the CATM model relevant to our discussions \cite{a4}. More details can also be found in \cite{bueno}. Consider the zero curvature condition $\partial
_{+}A_{-}-\partial _{-}A_{+}+[A_{+},A_{-}]=0$. The potentials take the
form
\br
A_{+}=-BF^{+}B^{-1},\quad A_{-}=-\partial _{-}BB^{-1}+F^{-},\qquad 
\er
with
\br
F^{+} \,=\,E^{3}+F_{1}^{+}+F_{2}^{+},\,\,\,\,\,\,
F^{-} \,=\, E^{-3}+F_{1}^{-}+F_{2}^{-},
\er
where $E^{\pm 3}\equiv {\bf m}.H^{\pm} =\frac{1}{6}[(2m^{1}_{\psi}+m^{2}_{\psi})H^{\pm 1}_{1}+(2m^{2}_{\psi}+m^{1}_{\psi})H^{\pm 1}_{2}]$ and the $F_{i}^{\pm }$'s and $B$ contain the spinor fields and scalars of the model, respectively 
\br
\lab{F1}
F_{1}^{+}&=&\sqrt{im^{1}_{\psi}}\psi _{R}^{1}E_{\alpha _{1}}^{0}+\sqrt{im^{2}_{\psi}}\psi
_{R}^{2}E_{\alpha _{2}}^{0}+\sqrt{im^{3}_{\psi}}\widetilde{\psi }_{R}^{3}E_{-\alpha
_{3}}^{1},
\\
\lab{F2}
F_{2}^{+}&=&\sqrt{im^{3}_{\psi}}\psi _{R}^{3}E_{\alpha _{3}}^{0}+\sqrt{im^{1}_{\psi}}
\widetilde{\psi }_{R}^{1}E_{-\alpha _{1}}^{1}+\sqrt{im^{2}_{\psi}}\widetilde{\psi }
_{R}^{2}E_{-\alpha _{2}}^{1},
\\
\lab{F3}
F_{1}^{-}&=&\sqrt{im^{3}_{\psi}}\psi _{L}^{3}E_{\alpha _{3}}^{-1}-\sqrt{im^{1}_{\psi}}
\widetilde{\psi }_{L}^{1}E_{-\alpha _{1}}^{0}-\sqrt{im^{2}_{\psi}}\widetilde{\psi }
_{L}^{2}E_{-\alpha _{2}}^{0},
\\
\lab{F4}
F_{2}^{-}&=&\sqrt{im^{1}_{\psi}}\psi _{L}^{1}E_{\alpha _{1}}^{-1}+\sqrt{im^{2}_{\psi}}\psi
_{L}^{2}E_{\alpha _{2}}^{-1}-\sqrt{im^{3}_{\psi}}\widetilde{\psi }
_{L}^{3}E_{-\alpha _{3}}^{0},\\
B&=&e^{i\vp_{1} H^{0}_{1}+i\vp_{2} H^{0}_{2} }\,e^{\widetilde{\nu }C}\,e^{\eta Q_{ppal}}, \lab{eq1}
\er
where $E_{\alpha _{i}}^{n},H^{n}_{1},H^{n}_{2}$ and  $C$ ($i=1,2,3; \, n=0,\pm 1$) are some generators of $sl(3)^{(1)}$; $Q_{ppal}$ being the principal gradation operator. The commutation relations for an affine Lie algebra in the Chevalley basis are
\br
&&\left[ \emph{H}_a^m,\emph{H}_b^n\right] =mC\frac{2}{\alpha_{a}^2}K_{a b}\delta _{m+n,0}  \lab{a7}\\
&&\left[ \emph{H}_a^m,E_{\pm \alpha}^n\right] = \pm K_{\alpha a}E_{\pm \alpha}^{m+n} 
\lab{a8}\\
&&\left[ E_\alpha ^m,E_{-\alpha }^n\right]
=\sum_{a=1}^rl_a^\alpha \emph{H}_a^{m+n}+\frac 2{\alpha ^2}mC\delta
_{m+n,0}  \lab{a9}
\\
&&\left[ E_\alpha ^m,E_\beta ^n\right] =
\varepsilon (\alpha ,\beta )E_{\alpha +\beta }^{m+n};\qquad \mbox{if }\alpha
+\beta \mbox{ is a root \qquad }  \lab{a10}
\\
&&\left[ D,E_\alpha ^n\right] =nE_\alpha ^n,\qquad \left[ D,\emph{H}%
_a^n\right] =n\emph{H}_a^n.  \lab{a12}
\er
where $K_{\alpha a}=2\a.\a_{a}/\a_{a}^2=n_{b}^{\a}K_{ba}$, with $n_{a}^{\a}$ and $l_a^\alpha$ being the integers in the expansions $\a=n_{a}^{\a}\a_{a}$ and $\a/\a^2=l_a^\alpha\a_{a}/\a_{a}^2$, and $\varepsilon (\alpha ,\beta )$ the relevant structure constants.  

Take $K_{11}=K_{22}=2$ and $K_{12}=K_{21}=-1$ as the Cartan matrix elements of the simple Lie algebra $sl(3)$. Denoting by $\a_{1}$ and $\a_{2}$ the simple roots and the highest one by $\psi (=\a_{1}+\a_{2})$, one has $l_{a}^{\psi}=1(a=1,2)$, and $K_{\psi 1}=K_{\psi 2}=1$. Take $\varepsilon (\alpha ,\beta )=-\varepsilon (-\alpha ,-\beta ),\,\, \varepsilon_{1,2}\equiv \varepsilon (\alpha_{1} ,\a_{2})=1,\,\,  \varepsilon_{-1,3}\equiv \varepsilon(-\alpha_{1} ,\psi )=1\,\, \mbox{and}\, \,\,\varepsilon_{-2,3}\equiv \varepsilon (-\alpha_{2} ,\psi)=-1$.

One has $Q_{ppal} \equiv \sum_{a=1}^{2}  {\bf s}_{a}\l^{v}_{a}.H + 3 D$, where $\l^{v}_{a}$ are the fundamental co-weights of $sl(3)$, and the principal gradation vector is ${\bf s}=(1,1,1)$ \cite{kac}. 

The zero curvature condition gives the following equations of motion 
\br
\lab{eqnm1}
\frac{\partial ^{2}\varphi_{a}}{4i\,e^{\eta}} &=&m^{1}_{\psi}[e^{\eta -i\phi_{a}}\widetilde{\psi }_{R}^{l}\psi _{L}^{l}+e^{i\phi_{a}}\widetilde{\psi }_{L}^{l}\psi
_{R}^{l}]+m^{3}_{\psi}[e^{-i\phi_{3}}\widetilde{\psi }
_{R}^{3}\psi _{L}^{3}+e^{\eta +i\phi_{3}}\widetilde{\psi }
_{L}^{3}\psi _{R}^{3}];\,\,\,\,a=1,2\,\,\,\,\,\,\,\,\,\,\,\,\,\,\,\,\,\,\,
\\
\lab{eqnm3}
-\frac{\partial ^{2}\widetilde{\nu }}{4} &=&im^{1}_{\psi}e^{2\eta -\phi_{1}}\widetilde{\psi }_{R}^{1}\psi _{L}^{1}+im^{2}_{\psi}e^{2\eta
-\phi_{2}}\widetilde{\psi }_{R}^{2}\psi
_{L}^{2}+im^{3}_{\psi}e^{\eta -\phi_{3}}\widetilde{\psi }
_{R}^{3}\psi _{L}^{3}+{\bf m}^{2}e^{3\eta },\,\,
\\
\lab{eqnm4}
-2\partial _{+}\psi _{L}^{1}&=&m^{1}_{\psi}e^{\eta +i\phi_{1}}\psi
_{R}^{1},\,\,\,\,\,\,\,\,\,\,\,\,\,\,\,
-2\partial _{+}\psi _{L}^{2}\,=\,m^{2}_{\psi}e^{\eta +i\phi_{2}}\psi
_{R}^{2},
\\
\lab{eqnm5}
2\partial _{-}\psi _{R}^{1}&=&m^{1}_{\psi}e^{2\eta -i\phi_{1}}\psi
_{L}^{1}+2i \(\frac{m^{2}_{\psi}m^{3}_{\psi}}{im^{1}_{\psi}}\)^{1/2}e^{\eta }(-\psi _{R}^{3}
\widetilde{\psi }_{L}^{2}e^{i\phi _{2}}-
\widetilde{\psi }_{R}^{2}\psi _{L}^{3}e^{-i\phi_{3}}),
\\
\lab{eqnm7}
2\partial _{-}\psi _{R}^{2}&=&m^{2}_{\psi}e^{2\eta -i\phi_{2}}\psi
_{L}^{2}+2i\(\frac{m^{1}_{\psi}m^{3}_{\psi}}{im^{2}_{\psi}}\)^{1/2}e^{\eta }(\psi _{R}^{3}
\widetilde{\psi }_{L}^{1}e^{i\phi _{1}}+
\widetilde{\psi }_{R}^{1}\psi _{L}^{3}e^{-i\phi _{3}}), 
\\
\lab{eqnm8}
-2\partial _{+}\psi _{L}^{3}&=&m^{3}_{\psi}e^{2\eta +i\phi _{3}}\psi
_{R}^{3}+2i\(\frac{m^{1}_{\psi}m^{2}_{\psi}}{im^{3}_{\psi}}\)^{1/2}e^{\eta }(-\psi _{L}^{1}\psi
_{R}^{2}e^{i\phi_{2}}+\psi _{L}^{2}\psi
_{R}^{1}e^{i\phi _{1}}),
\\
\lab{eqnm9}
2\partial _{-}\psi _{R}^{3}&=&m^{3}_{\psi}e^{\eta -i\phi_{3}}\psi
_{L}^{3},\,\,\,\,\,\,\,\,\,\,\,\,
2\partial _{-}\widetilde{\psi }_{R}^{1}\,=\,m^{1}_{\psi}e^{\eta +i\phi_{1}}\widetilde{\psi }_{L}^{1},
\\
\lab{eqnm10}
-2\partial _{+}\widetilde{\psi }_{L}^{1} &=&m^{1}_{\psi}e^{2\eta -i\phi_{1}}\widetilde{\psi }_{R}^{1}+2i\(\frac{m^{2}_{\psi}m^{3}_{\psi}}{im^{1}_{\psi}
}\)^{1/2}e^{\eta }(-\psi _{L}^{2}\widetilde{\psi }_{R}^{3}e^{-i\phi _{3}}-\widetilde{\psi }_{L}^{3}\psi _{R}^{2}e^{i\phi_{2}}), 
\\
\lab{eqnm12}
-2\partial _{+}\widetilde{\psi }_{L}^{2} &=&m^{2}_{\psi}e^{2\eta -i\phi_{2}}\widetilde{\psi }_{R}^{2}+2i\(\frac{m^{1}_{\psi}m^{3}_{\psi}}{im^{2}_{\psi}}
\)^{1/2}e^{\eta }(\psi _{L}^{1}\widetilde{\psi }_{R}^{3}e^{-i\phi _{3}}+\widetilde{\psi }_{L}^{3}\psi _{R}^{1}e^{i\phi_{1}}), 
\\
\lab{eqnm13}
2\partial _{-}\widetilde{\psi }_{R}^{2}&=&m^{2}_{\psi}e^{\eta+i\phi_{2}}\widetilde{\psi }_{L}^{2}, \,\,\,\,\,\,\,\,\,\,\,\,\,\,\,\,
-2\partial _{+}\widetilde{\psi }_{L}^{3}\,=\,m^{3}_{\psi}e^{\eta -i\phi _{3}}\widetilde{\psi }_{R}^{3}, 
\\
\lab{eqnm15}
2\partial _{-}\widetilde{\psi }_{R}^{3} &=&m^{3}_{\psi}e^{2\eta +i\phi_{3}}\widetilde{\psi }_{L}^{3}+2i\(\frac{m^{1}_{\psi}m^{2}_{\psi}}{im^{3}_{\psi}}
\)^{1/2}e^{\eta }(\widetilde{\psi} _{R}^{1}\widetilde{\psi }_{L}^{2}e^{i\phi_{2}}-\widetilde{\psi }_{R}^{2}\widetilde{\psi }_{L}^{1}e^{i\phi_{1}}), 
\\
\lab{eqnm16}
\partial^{2}\eta&=&0,
\er
where $\phi_{1}\equiv2 \vp_{1}-\vp_{2},\,\phi_{2}\equiv 2\vp_{2}-\vp_{1},\,\phi_{3} \equiv \vp_{1}+\vp_{2}$.

Apart from the {\sl conformal invariance} the above equations exhibit the $\(U(1)_{L}\)^{2}\otimes \(U(1)_{R}\)^{2}$ {\sl left-right local gauge symmetry}
\br
\lab{leri1}
\vp_{a} &\ra& \vp_{a} + \theta_{+}^{a}( x_{+}) + \theta_{-}^{a}( x_{-}),\,\,\,\,a=1,2\\
\widetilde{\nu} &\ra& \widetilde{\nu}\; ; \qquad \eta \ra \eta \\
\psi^{i} &\ra & e^{i( 1+ \gamma_5) \Theta_{+}^{i}( x_{+}) 
+ i( 1- \gamma_5) \Theta_{-}^{i}( x_{-})}\, \psi^{i},\\
\,\,\,\,
\widetilde{\psi}^{i} &\ra& e^{-i( 1+ \gamma_5) (\Theta_{+}^{i})( x_{+})-i ( 1- \gamma_5) (\Theta_{-}^{i})( x_{-})}\,\widetilde{\psi}^{i},\,\,\, i=1,2,3;\lab{leri2}
\\
&&\Theta^{1}_{\pm}\equiv \pm \theta_{\pm}^{2} \mp 2\theta_{\pm}^{1},\,\,\Theta^{2}_{\pm}\equiv \pm \theta_{\pm}^{1}\mp 2\theta_{\pm}^{2},\,\,\Theta_{\pm}^{3}\equiv \Theta_{\pm}^{1}+\Theta_{\pm}^{2}.\nn
\er

One can get global symmetries for $\theta_{\pm}^{a}=\mp \theta_{\mp}^{a}=$ constants. For a model defined by a Lagrangian these would imply the presence of two vector and two chiral conserved currents. However, it was found only half of such currents \cite{bueno}. This is a consequence of the lack of a Lagrangian description for the $sl(3)^{(1)}$ CATM; however see below. 

The gauge fixing of the conformal symmetry, by setting the field $\eta$ to a constant, is used to stablish the $U(1)$ vector, $J^{\mu}=\sum_{j=1}^{3} m^{j}_{\psi} \bar{\psi}^{j}\gamma^{\mu}\psi^{j}$, and topological currents equivalence \cite{matter,annals}. Moreover, it has been shown that the soliton solutions are in the orbit of the solution $\eta=0$. The remarkable equivalence is
\br
\lab{equivalence}
\sum_{j=1}^{3} m^{j}_{\psi} \bar{\psi}^{j}\gamma^{\mu}\psi^{j} \equiv \epsilon^{\mu \nu}\partial_{\nu} (m^{1}_{\psi}\varphi_{1}+m^{2}_{\psi}\varphi_{2}),\,\,\,\,\,\,\, m^{3}_{\psi}=m^{1}_{\psi}+ m^{2}_{\psi},\,\,\,\,m^{i}_{\psi}>0.
\er

The CATM theory has a local Lagrangian in terms of the $B$ and the (two-loop) WZNW fields \cite{matter}. The relations between their fields can be obtained from\br
\lab{def11}
F^{-}=B\partial _{-}MM^{-1}B^{-1}\,\,\,\mbox{ and }\,\,\,
F^{+}=B^{-1}N^{-1}\partial _{+}NB 
\er
where 
\br
\lab{def22}
M=\exp (\sum_{s>0}\zeta _{s}),\,\,\,\,\,\,N=\exp (\sum_{s>0}\xi _{-s}), 
\er
provided that the following constraints are imposed  
\br
\lab{const1}
(\partial _{-}MM^{-1})_{-3}=B^{-1}({\bf m}.H^{-1})B,\quad (\partial
_{-}MM^{-1})_{<-3}=0. 
\er
and 
\br
\lab{const2}
(N^{-1}\partial_{+}N)_{3}=B\,({\bf m}.H^{1})B^{-1},\,\,\,\, (N^{-1}\partial
_{+}N)_{>3}=0. 
\er

In \rf{def22} and \rf{const1}-\rf{const2} $s$ and the subscripts denote the principal gradation structure of the relevant group elements. 

The relationships are
\br
\nn
\sqrt{im^{3}_{\psi}}\psi _{L}^{3}&=&e^{-\eta +i\phi_{3}}\partial
_{-}\xi _{-1}^{3},\,\,\,\,\,\,
-\sqrt{im^{1}_{\psi}}\widetilde{\psi }_{L}^{1}=e^{-\eta -i\phi_{1}}\partial _{-}\xi _{-1}^{1},\\
\nn
-\sqrt{im^{2}_{\psi}}\widetilde{\psi }_{L}^{2}&=&e^{-\eta -i\phi_{2}}\partial _{-}\xi _{-1}^{2},\,\,\,\,\,\,\,\,\,\,\,
\sqrt{im^{3}_{\psi}}\psi _{R}^{1}=e^{\eta -i\phi_{1}}\partial
_{+}\zeta _{1}^{1},\\
\nn
\sqrt{im^{2}_{\psi}}\psi _{R}^{2}&=&e^{\eta -i\phi_{2}}\partial
_{+}\zeta _{1}^{2},\,\,\,\,\,\,\,\,\,\,\,\,\,\,\,\,\,
\sqrt{im^{3}_{\psi}}\widetilde{\psi }_{R}^{3}\,=\,e^{\eta +i\phi_{3}}\partial _{+}\zeta _{1}^{3},
\\
\nn
\sqrt{im^{1}_{\psi}}\psi _{L}^{1} e^{2\eta} &=&e^{i\phi_{1}}\partial_{-}\xi _{-2}^{1}+\frac{1}{2} (\xi _{-1}^{3}\partial _{-}\xi_{-1}^{2}\epsilon _{3,-2}+\xi _{-1}^{2}\partial _{-}\xi _{-1}^{3}\epsilon
_{-2,3})e^{i\phi_{1}},
\\
\lab{relationships}
\sqrt{im^{2}_{\psi}}\psi _{L}^{2} e^{2\eta} &=&e^{i\phi_{2}}\partial
_{-}\xi _{-2}^{2}+\frac{1}{2} (\xi _{-1}^{3}\partial _{-}\xi
_{-1}^{1}\epsilon _{3,-1}+\xi _{-1}^{1}\partial _{-}\xi _{-1}^{3}\epsilon
_{-1,2})e^{i\phi_{2}},
\\
\nn
-\sqrt{im^{3}_{\psi}}\widetilde{\psi }_{L}^{3} e^{2\eta}&=&e^{-i\phi _{3}}\partial _{-}\xi _{-2}^{3}+\frac{1}{2} (\xi _{-1}^{1}\partial
_{-}\xi _{-1}^{2}\epsilon _{-1,-2}+\xi _{-1}^{2}\partial _{-}\xi
_{-1}^{1}\epsilon _{-2,-1})e^{-i\phi _{3}},
\\
\nn
\sqrt{im^{3}_{\psi}}\psi _{R}^{3} e^{-2\eta}&=&e^{-i\phi_{3}}\partial
_{+}\zeta _{2}^{3}-\frac{1}{2} (\zeta _{1}^{1}\partial _{+}\zeta
_{1}^{2}\epsilon _{1,2}+\zeta _{1}^{2}\partial _{+}\zeta _{1}^{1}\epsilon
_{2,1})e^{-i\phi_{3}},
\\
\nn
\sqrt{im^{1}_{\psi}}\widetilde{\psi }_{R}^{1} e^{-2\eta}&=&e^{i\phi_{1}}\partial _{+}\,\zeta _{2}^{1}-\frac{1}{2} (\zeta
_{1}^{3}\partial _{+}\zeta _{1}^{2}\epsilon _{-3,2}+\zeta _{1}^{2}\partial
_{+}\zeta _{1}^{3}\epsilon _{2,-3})e^{i\phi_{1}},
\\
\nn
\sqrt{im^{2}_{\psi}}\widetilde{\psi }_{R}^{2} e^{-2\eta}&=&e^{i\phi_{2}}\partial _{+}\zeta _{2}^{2}-\frac{1}{2} (\zeta
_{1}^{3}\partial _{+}\zeta _{1}^{1}\epsilon _{-3,1}+\zeta _{1}^{1}\partial
_{+}\zeta _{1}^{3}\epsilon _{1,-3})e^{i\phi_{2}}.
\er

We observe that the WZNW fields $\xi_{-1}^{i}$,\,$\xi_{-2}^{i}$,\,$\zeta_{1}^{i}$,\, $\zeta_{2}^{i}$\, ($i=1,2,3$) are nonlocal in terms of the spinors and scalars \{$\psi_{i}, \widetilde{\psi}_{i}, \varphi_{1}, \varphi_{2}, \widetilde{\nu}\,\, \mbox{and}\,\, \eta $\}. Then the CATM model Lagrangian must be nonlocal when written in terms of its fields.

\end{document}